\documentclass[useAMS,usenatbib]{mn2e}
\usepackage[T1]{fontenc} 
\usepackage{aecompl}
\usepackage{pdflscape}
\usepackage{longtable}
\usepackage[para]{threeparttable}
\usepackage[normalem]{ulem}
\usepackage{lmodern}

\newcommand{\hii}{\mbox{H~{\sc ii}~}}
\newcommand{\mgi}{\mbox{Mg~{\sc i}~}}
\newcommand{\mgii}{\mbox{Mg~{\sc ii}~}}
\newcommand{\mgiii}{\mbox{Mg~{\sc iii}~}}
\newcommand{\oii}{\mbox{O~{\sc ii}~}}

\newcommand{\siii}{\mbox{Si~{\sc ii}~}}
\newcommand{\siiii}{\mbox{Si~{\sc iii}~}}
\newcommand{\siiv}{\mbox{Si~{\sc iv}~}}
\newcommand{\ciii}{\mbox{C~{\sc iii}~}}
\newcommand{\heii}{\mbox{He~{\sc ii}~}}
\newcommand{\hei}{\mbox{He~{\sc i}~}}
\newcommand{\nai}{\mbox{Na~{\sc i}~}}
\newcommand{\cai}{\mbox{Ca~{\sc i}~}}
\newcommand{\feii}{\mbox{Fe~{\sc ii}~}}

\title[The young cluster NGC 2282 : a multi-wavelength perspective]{The young cluster NGC 2282 : a multi-wavelength perspective}
\author[S. Dutta, S. Mondal $\&$ J. Jose   et al.]{Somnath Dutta$^{1}$\thanks{E-mail: somnath12@boson.bose.res.in (SD)},  S. Mondal$^{1}$, J. Jose$^{2}$,  R. K. Das$^{1}$, M. R. Samal$^{3}$,  and S. Ghosh$^{1}$\\
$^{1}$S.N. Bose National Centre for Basic Sciences, Kolkata 700098, India\\
$^{2}$Kavli Institute for Astronomy and Astrophysics,Peking University,
Yi He Yuan Lu 5, Haidian District, Beijing 100871, China\\
$^{3}$Aix Marseille Universit\'{e}, CNRS, LAM (Laboratoire d'Astrophysique de Marseille) UMR 7326, 13388 Marseille, France}

\usepackage{graphicx}
\begin{document}

\date{Accepted 2... December 00. Received 2... December 00; in original form 2.... October ..}

\pagerange{\pageref{firstpage}--\pageref{lastpage}} \pubyear{2015}

\maketitle

\label{firstpage}

\begin{abstract}

We present the analysis of the stellar content of NGC~2282, a young cluster in the Monoceros constellation, using deep optical $BVI$ and IPHAS photometry along with infrared (IR) data from UKIDSS  and $Spitzer$-IRAC. Based on the stellar surface density analysis using nearest neighborhood method, the radius of the cluster is estimated as $\sim$ 3.15$\arcmin$. From optical spectroscopic analysis of 8 bright sources, we have  classified three early B-type members in the cluster, which includes,  HD 289120, a previously known  B2V type star,  a Herbig Ae/Be star (B0.5 Ve) and a B5 V star. From spectrophotometric analyses, the distance to the cluster has been estimated as $\sim$ 1.65 kpc. The $K$-band extinction map is estimated using nearest neighborhood technique, and the mean extinction within the cluster area is found to be A$_V$ $\sim$ 3.9 mag.  Using IR colour-colour criteria and H$_\alpha$-emission properties, we have identified a total of 152 candidate young stellar objects (YSOs) in the region, of which, 75 are classified as Class II, 9 are Class I YSOs. Our YSO catalog also includes 50 H$_\alpha$-emission line sources, identified using slitless spectroscopy and IPHAS photometry data.  Based on the optical and near-IR colour-magnitude diagram analyses, the  cluster age has been estimated to be in the range of 2 $-$ 5 Myr, which is in agreement with the estimated age from disc fraction ($\sim$ 58\%). Masses of these YSOs are found to be $\sim$ 0.1$-$2.0 M$_\odot$.  Spatial distribution of the candidate YSOs shows spherical morphology, more or less similar to the surface density map.  

\end{abstract}

\begin{keywords}
Embedded clusters---young stellar objects -- infrared: stars.
\end{keywords}

\section{Introduction}

 Young embedded stellar clusters are an active star-forming site, where pre-main sequence (PMS) massive and low-mass stars are formed together, and evolved over timescale of few million years (Lada et al. 1994; Lada \& Lada 2003). Multiwavelength studies of these young stellar objects (YSOs) provide unique opportunity to understand their formation, evolution over time, circumstellar discs including planet formation in their discs, environment induced formation scenario and history (e.g. Carpenter et al. 2001; Kenyon S., \& Hartman 1995; Evans et al. 2009; Lada et al. 2010; Jose et al. 2012, 2013; Samal et a. 2015). Highly uncertain issue of contamination from foreground and background population could be minimized from YSOs excesses at near-IR and mid-IR wavelengths and observations of close reference field at similar depth (Allen et al. 2004; Megeath et al. 2004; Flaherty et a. 2007). Such contamination could be further narrow down through spectroscopic observations in the optical and near-IR wavelengths (Hillenbrand 1997; Briceno et al. 2002; Luhman 2004; Rebull et al. 2010; Herczeg \& Hillenbrand 2014).

\begin{figure*}
\includegraphics[width=8 cm,height=8.0cm,bb=130 225 460 550 ]{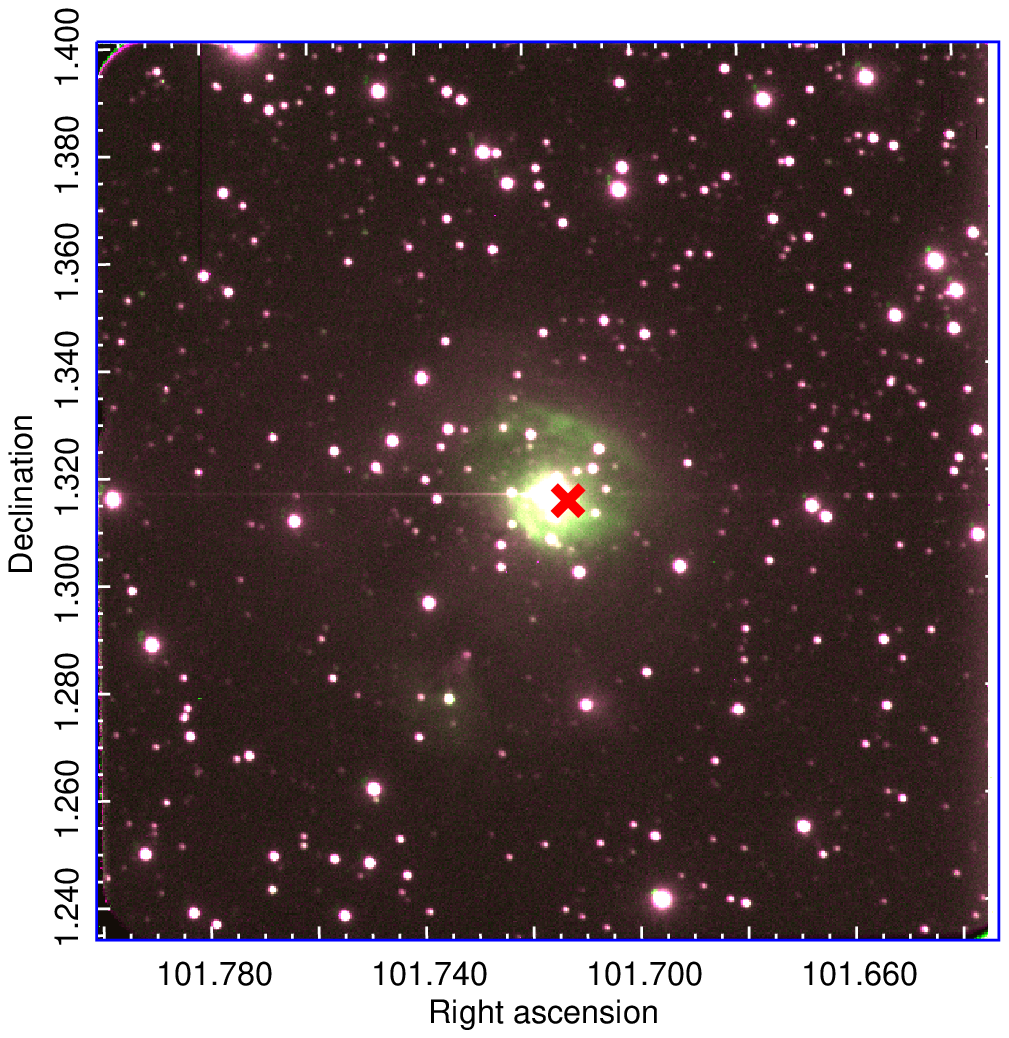}
\includegraphics[width=8 cm,height=8.0cm,bb=60 130 530 630]{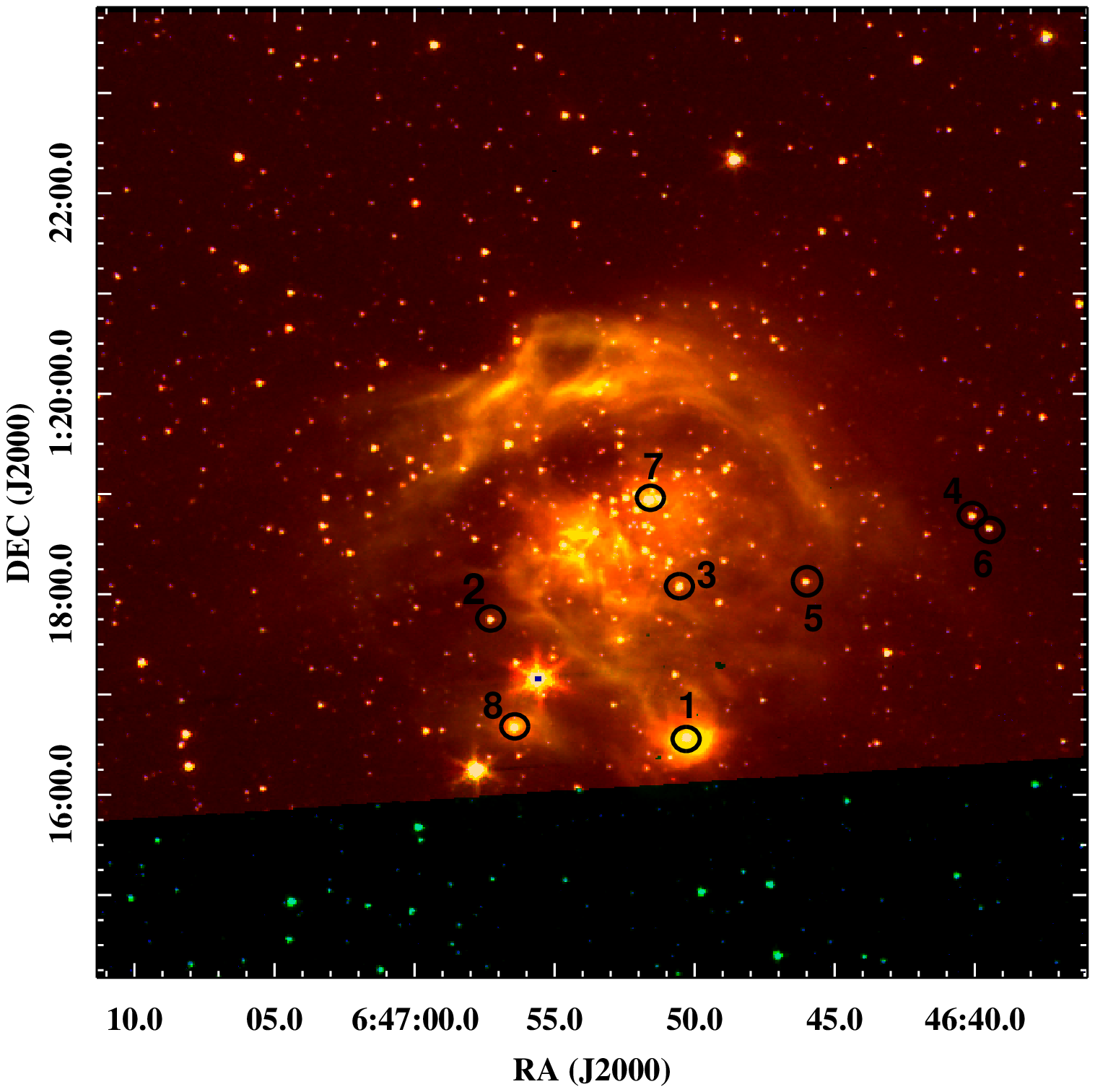}
  \caption{(a) Optical colour composite image of NGC 2282 (blue: 5007 \AA (O III); green: 6563 \AA (H$_\alpha$); red: 6724 \AA [S II]) obtained using 2m HCT. The cluster centre is marked with red cross. (b) Near-IR Colour composite image of NGC 2282 (blue: $K$ band; green: 3.6 $\mu$m; red: 4.5 $\mu$m). All the spectroscopically studied objects are numbered with black circles (see text for details).}
  \label{fig:optical_nir}
\end{figure*}

\begin{table*}
\small
\centering
 \begin{minipage}{140mm}
  \caption{Log of Observations.} 
\label{tab:observation} 
\begin{tabular}{|c|c|c|c|c|c|c|c|c|}
\hline

ID & $\alpha_{(2000)}$ & $\delta_{(2000)}$ & Date of& Grism/filter & Exp. time (s) & Airmass & SNR \\
  & (h:m:s)&(d:m:s)& Observations &  &$\times$ no. frame & & \\
\hline
\hline
\multicolumn{8}{c}{optical photometry}\\
\hline
NGC 2282 &06:46:50.4&+01:18:50&14.12.2007&B&600$\times$6, 30$\times$1 & 1.048 & \\
NGC 2282 &06:46:50.4&+01:18:50&14.12.2007&V&600$\times$6, 10$\times$1 & 1.242 & \\
NGC 2282 &06:46:50.4&+01:18:50&14.12.2007&I&300$\times$5, 10$\times$1 & 1.339 & \\
\hline
\multicolumn{8}{c}{slit spectroscopy}\\
\hline
 1  & 06:46:50.304 & +1:16:35.904 & 06.10.2014 &Gr7 &1500$\times$2 & 1.398 & 28 \\
 2  & 06:46:57.282 & +1:17:45.491 & 05.01.2013 &Gr7 &1800$\times$1 & 1.286 & 25 \\
 3  & 06:46:50.544 & +1:18:04.859 & 05.01.2013 &Gr7 &2500$\times$1 & 1.733 & 24 \\
 4  & 06:46:40.092 & +1:18:47.447 & 05.01.2013 &Gr7 &2500$\times$1 & 1.212 & 25 \\
 5 & 06:46:45.986 & +1:18:07.775 & 05.01.2013 &Gr7 &1500$\times$1 & 1.733 & 30 \\
 6 & 06:46:39.456 & +1:18:38.500 & 05.01.2013 &Gr7 &1500$\times$1 & 1.212 & 36 \\
 7 & 06:46:51.581 & +1:18:57.756 & 05.01.2013 &Gr7 &1500$\times$1 & 1.172 & 60 \\
 8  & 06:46:56.424 & +1:16:40.944 & 05.01.2013 &Gr7 &2500$\times$1 & 2.250 & 18 \\
\hline
\multicolumn{8}{c}{slitless spectroscopy}\\
\hline
NGC 2282   &06:46:49.4 &+1:18:44.5&08.11.2013&Gr8/H$_\alpha$-Br&1800$\times$1 & 1.181 & \\
NGC 2282   &06:46:49.4 &+1:18:44.5&25.01.2011&Gr5/H$_\alpha$-Br&1200$\times$1 & 1.532 & \\
\hline\end{tabular}
\end{minipage}
\end{table*}

We present here multiwavelength studies of the embedded cluster NGC~2282 ($\alpha_{2000}$ = $06^h46^m50.4^s$ $\delta_{2000}$ = $+01^018^m50^s$ ), a reflection nebula in the Monoceros constellation. It is located in an isolated molecular cloud of few thousand solar masses (Horner et al. 1997). The cluster is about 3$^o$ away on the sky from Mon OB2 and is probably associated with it. NGC~2282  has been listed in the several sky surveys of reflection nebulae (Van der Bergh 1966; Racine 1968; Kutner et al. 1980; Chini et al. 1984), and is also listed in the surveys of Galactic \hii regions as  BFS 54 (Blitz et al. 1982; Avedisova \& Kondratenko 1984; Fich 1993; Kislyakov \& Turner 1995). The distance of  NGC~2282 was estimated to be 1.7 $\pm$ 0.4 kpc, which was mainly based on the spectral type (B2 V) of the brightest star HD~289120 in NGC~2282 (Racine 1968; Avedisova \& Kondratenko 1984). So far three optical sources  were classified towards NGC~2282, namely V507 Mon, HD~289120 and EQ~0644.3+0121, of which, V507 Mon was identified as a spectroscopic binary (Wachmann 1996), while HD~289120 as the illuminating source to the reflection nebula (Horner et al. 1997). EQ~0644.3+0121 was noticed as a faint nebulous source in the Palomar plates (Petrossian 1985).

The cluster properties of NGC~2282 were first studied using $JHK$-bands data by Horner et al. (1997). The authors estimated the cluster age as $\sim$ 5$-$10 Myrs based on  the fraction (9\%) of infrared excess emission sources, and its association with the parent molecular cloud.  Horner et al. (1997) found a core radius 0.19 pc and a cluster radius 1.6 pc at 1.7 kpc distance, by radial profile fitting of $K$-band data with different models. Following the method of Lada et al. (1994), the authors further estimated a variable interstellar extinction in the cluster region, which showed a maximum  value of  $A_V$ $\sim$  6.7 $\pm$ 0.4 mag towards South-East part of the cluster centre, while quite low value of 1.6 $\pm$ 0.4 mag at the cluster centre around HD 289120.   

However, the fundamental parameters of the cluster are still not clearly understood; requires proper identification and 
characterization of the PMS sources present in the region. Due to the absence of spectroscopic or longer wavelength (e.g.,  disc and
envelope sources seen at  $\lambda$ $>$ 2.2 $\mu$m) or accretion signature indicator (e.g, emission line stars seen with H$_\alpha$ emission) observations, the identification and evolutionary status of the PMS sources in the region could not be studied so far.
Now with observations from Spitzer Space Telescope, in combination with deep $JHK$ data from UKDISS survey and $H_\alpha$ photometric data from IPHAS survey,  it is possible to make a more reliable membership census of the cluster by identifying its young stars with accretion discs and cold envelopes. In conjugation with  deep optical photometric data the identified PMS can be better characterized (e.g., mass, age)  by comparing their positions on the HR digram with the theoretical evolutionary models.  The characterization PMS stars are essential to derive the cluster properties and  to understand its star formation history (e.g., Lada \& Lada 1991; Persi et al. 1994; Tapia et al. 1997; Ojha et al. 2004).  We therefore studied the cluster with multiwavelength photometric data to have a better picture  on the cluster properties and  star formation activity.    

In this paper, we have estimated the basic properties of the cluster using deep optical/IR imaging and optical spectroscopic observations. We have identified and classified YSOs  based on their IR excess emission  in near-IR and mid-IR data  as well as their $H_\alpha$ emission line properties. We have characterized  the YSOs based on various colour-magnitude diagrams. Optical/IR colour composite image of NGC 2282 is shown in Fig.~\ref{fig:optical_nir}. These optical observations were taken by us using narrow-band filters at 5007 \AA [O III] (blue),  6563 \AA [$H_\alpha$] (green) and 6724 \AA [S II]) (red). While IR colour composite image is made using UKIDSS $K$-band (blue), {\it Spitzer} 3.6 $\mu$m (green) and 4.5 $\mu$m (red) images. Section~2 describes our observations and archival data sets used for the present study. Section~3 deals with our analyses and results that include spectroscopic study of 8 optical bright objects, identification and classification of the YSOs. In Section~4, we discuss the cluster properties such as ages, masses, spatial distribution etc. In section~5, we summarize our main results of the work.

\section[]{Data Sets Used}
\subsection{Observations}
\subsubsection{Optical Photometry} 
The CCD {\it BVI} observations of the cluster were acquired using the 1.04m Sampurnanand telescope (ST; Sagar 1999) at  Nainital, India during 2007 December 14. The log of optical observations is shown in Table~\ref{tab:observation}. We used a $2048 \times 2048$ CCD Camera having a pixel size of  24 $\mu$m and  field-of-view (FOV) about 13$\arcmin \times 13\arcmin$ with a plate scale of 0.37 arcsec pixel$^{-1}$. The gain and read out noise of the CCD are 10 e{$^-$}/Analog to Digital Unit(ADU) and  5.3 e$^{-}$ respectively. The observations were taken in 2$\times$2 binning mode to improve the signal to noise ratio (SNR) and the average FWHM of the stars were $\sim$ 2$\arcsec$. The observations were taken in short and long exposures to get a good dynamic coverage of the stellar brightness. Along with the NGC 2282 field, standard stars in the SA 92 field (Landolt 1992) were also observed at various airmasses on the same night.

\begin{figure}

\includegraphics[width=8.0 cm,height=8.5cm, angle=0]{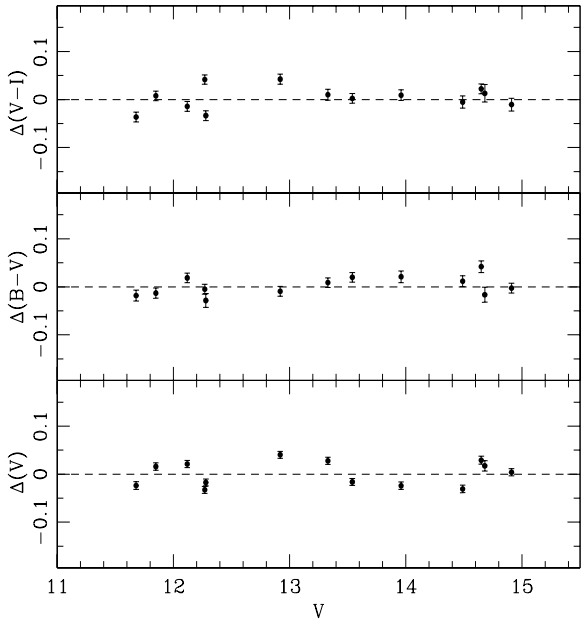}
  \caption{Residuals between standard and transformed magnitudes and colours of standard stars plotted against the Landolt standard magnitudes. The error bars are combined errors of Landolt (1992) and present measurements. }  
  \label{fig:residual}  
\end{figure}

The raw CCD images were cleaned using IRAF\footnote{Image Reduction and Analysys Facility (IRAF) is distributed by National Optical Astronomy Observatories (NOAO), USA (http://iraf.noao.edu/)} software following bias subtraction, flat fielding and cosmic ray removal. The identification of point sources was performed with the DAOFIND task. Following Stetson (1987), we have used the roundness limits of $-1$ to $+1$ and sharpness limits of $0.2$ to $+1$  to eliminate bad pixels  brightness enhancements and the extended sources such as background galaxies from the point source catalog. The photometry by PSF fitting was done using ALLSTAR task of DAOPHOT package (Stetson 1992).  The instrumental magnitudes were converted to standard magnitudes following the procedure outlined by Stetson (1987). A total of 13 stars of SA 92 field were used to estimate atmospheric extinction and transformation coefficients. The estimated extinction coefficients in $B$, $V$ and $I$ are 0.307 $\pm$ 0.014, 0.185 $\pm$ 0.005 and 0.108 $\pm$ 0.008, respectively. The final transformation equations used for photometric calibrations are

\begin{equation}
(V-I) = (0.971\pm 0.008)(v-i) + (0.511\pm0.006)
\end{equation}
\begin{equation}
(B-V) = (0.977\pm 0.009)(b-v)  - (0.29 \pm 0.010)
\end{equation}
\begin{equation}
V = v + (0.089 \pm 0.001)(V-I) - (4.31\pm 0.002) 
\end{equation}

   where $B$, $V$, $I$ are the standard magnitudes and $b$, $v$, $i$ are the instrumental magnitudes corrected for the atmospheric extinctions for the airmass given in Table~\ref{tab:observation}. The error of final magnitude measurements are obtained by propagating the  uncertainties in extinction measurements, standard coefficients and  profile-fitting photometry etc. The profile-fitting uncertainty is estimated from  the gain and read-out-noise of CCD camera,  and background level of an image etc. Fig.~\ref{fig:residual} shows the standardization residuals ($\Delta$) between standard and transformed $V$ magnitudes, $(B - V)$ and $(V - I)$ colours of standard stars as function of $V$ magnitudes. The standard deviations in $\Delta V$, $\Delta (B - V)$ and $\Delta (V - I)$  are 0.024, 0.019 and 0.023 mag, respectively.  Finally, we estimated optical magnitudes of 1379 objects, which are detected in two or more bands with a limiting magnitude of $V$ $\sim$ 22 mag.

      The  world coordinate system (wcs) coordinates for  the  detected  stars  in  the  frame  were  determined  using  20 isolated moderately bright stars with their positions from the 2MASS point source catalogue (PSC) (Curti et al. 2003), and a position accuracy of better than 0.3$\arcsec$ has been achieved. We used IRAF tasks $ccfind$, $ccmap$ and $ccsetwcs$ to achieve the above astrometric solution. 
 
\subsubsection{Slit Spectroscopy}
We obtained optical spectroscopic observations of 8 bright sources within NGC 2282  using HFOSC of 2m Himalayan Chandra Telescope (HCT), India (Prabhu 2014). There observations were acquired using Grism 7 (3800- 6840 {\AA})  with  a resolving power of 1200 and a spectral dispersion of about 2.9 {\AA} per two pixels. The   FeAr arc lamp observations were taken immediately after the target observations. The spectrophotometric standard star G191B2B (Oke 1990) was also observed with an exposure time of 600s for the flux calibration. The log of spectroscopic observations is tabulated in Table~\ref{tab:observation}. 

  After bias subtraction and flat field correction, the one-dimensional spectra were extracted using the optimal extraction method using APALL task in IRAF and wavelength calibrated using the FeAr arc lamp observations. The spectra were also corrected for the instrumental response using the sensitivity function generated from the standard star observations.

\renewcommand{\tabcolsep}{3.0pt} 
\begin{table*}
\caption{Photometric catalog of point sources towards NGC 2282. The complete table is available in the electronic version.}
\tiny

\label{tab:catalog}

\begin{tabular}{rrrrrrrrrrrr}
\hline \multicolumn{1}{c}{ID} & \multicolumn{1}{c}{RA (J2000)} & \multicolumn{1}{c}{Dec (J2000)} & \multicolumn{1}{c}{$V$} & \multicolumn{1}{c}{$B-V$} & \multicolumn{1}{c}{$V-I$} & \multicolumn{1}{c}{$J$} & \multicolumn{1}{c}{$H$} & \multicolumn{1}{c}{$K$} & \multicolumn{1}{c}{3.6 $\mu$m} & \multicolumn{1}{c}{4.5 $\mu$m} \\ 
\multicolumn{1}{c}{} & \multicolumn{1}{c}{(deg)} & \multicolumn{1}{c}{(deg)} & \multicolumn{1}{c}{(mag)} & \multicolumn{1}{c}{(mag)} & \multicolumn{1}{c}{(mag)} & \multicolumn{1}{c}{(mag)} & \multicolumn{1}{c}{(mag)} & \multicolumn{1}{c}{(mag)} & \multicolumn{1}{c}{(mag)} & \multicolumn{1}{c}{(mag)} \\ \hline
\hline
    1  &  101.709618  &  1.276534  &  14.674 $\pm$  0.008  &   0.933  $\pm$  0.007  &    1.438  $\pm$  0.004 &   11.964  $\pm$  0.020  &  11.506  $\pm$  0.029  &  11.164  $\pm$   0.034  &  10.678  $\pm$  0.002  &  10.371  $\pm$  0.003\\
    2  &  101.738704  &  1.295886  &  13.858 $\pm$  0.006  &   0.775  $\pm$  0.003  &    0.945  $\pm$  0.002 &   12.273  $\pm$  0.019  &  11.886  $\pm$  0.022  &  11.771  $\pm$   0.023  &  11.706  $\pm$  0.004  &  11.712  $\pm$  0.005\\
    3  &  101.710595  &  1.301281  &  14.285 $\pm$  0.004  &   0.978  $\pm$  0.003  &    1.177  $\pm$  0.003 &   12.264  $\pm$  0.020  &  11.789  $\pm$  0.025  &  11.648  $\pm$   0.025  &  11.606  $\pm$  0.004  &  11.612  $\pm$  0.005\\
    4  &  101.667052  &  1.313098  &  13.348 $\pm$  0.005  &   0.587  $\pm$  0.002  &    0.735  $\pm$  0.005 &   12.196  $\pm$  0.020  &  11.959  $\pm$  0.026  &  11.873  $\pm$   0.026  &  11.816  $\pm$  0.004  &  11.824  $\pm$  0.005\\
    5  &  101.691841  &  1.302181  &  13.777 $\pm$  0.005  &   0.721  $\pm$  0.003  &    0.803  $\pm$  0.002 &   12.460  $\pm$  0.020  &  12.155  $\pm$  0.026  &  12.080  $\pm$   0.025  &  12.017  $\pm$  0.004  &  12.066  $\pm$  0.006\\
    6  &  101.664463  &  1.311054  &  14.037 $\pm$  0.004  &   0.646  $\pm$  0.003  &    0.809  $\pm$  0.005 &   12.711  $\pm$  0.020  &  12.548  $\pm$  0.027  &  12.429  $\pm$   0.028  &  12.333  $\pm$  0.005  &  12.318  $\pm$  0.006\\
    7  &  101.715141  &  1.315808  &  10.198 $\pm$  0.010  &   0.148  $\pm$  0.010  &    0.337  $\pm$  0.009 &    9.717  $\pm$  0.025  &   9.586  $\pm$  0.024  &   9.542  $\pm$   0.026  &   9.425  $\pm$  0.002  &   9.404  $\pm$  0.002\\
    8  &  101.735121  &  1.277934  &  15.934 $\pm$  0.013  &   1.037  $\pm$  0.010  &    1.841  $\pm$  0.007 &   12.061  $\pm$  0.023  &  11.212  $\pm$  0.020  &  10.572  $\pm$   0.023  &   9.616  $\pm$  0.003  &   9.049  $\pm$  0.002\\
    9  &  101.738965  &  1.208942  &  18.459 $\pm$  0.004  &   1.934  $\pm$  0.035  &    2.624  $\pm$  0.008 &   13.554  $\pm$  0.021  &  12.615  $\pm$  0.028  &  12.198  $\pm$   0.023  &  11.951  $\pm$  0.004  &         $...$   $~~~~~~$      \\
   10  &  101.662814  &  1.214660  &  16.779 $\pm$  0.002  &   1.506  $\pm$  0.010  &    1.914  $\pm$  0.010 &   13.243  $\pm$  0.020  &  12.477  $\pm$  0.027  &  12.242  $\pm$   0.026  &  12.064  $\pm$  0.004  &         $...$  $~~~~~~$       \\
   11  &  101.689942  &  1.216617  &  13.879 $\pm$  0.003  &   0.543  $\pm$  0.005  &    0.755  $\pm$  0.010 &   12.630  $\pm$  0.020  &  12.391  $\pm$  0.027  &  12.287  $\pm$   0.026  &  12.258  $\pm$  0.005  &         $...$   $~~~~~~$      \\
   12  &  101.705790  &  1.222968  &  15.948 $\pm$  0.003  &   0.921  $\pm$  0.004  &    1.286  $\pm$  0.007 &   13.620  $\pm$  0.023  &  13.225  $\pm$  0.027  &  13.064  $\pm$   0.033  &  12.890  $\pm$  0.007  &         $...$   $~~~~~~$      \\
   13  &  101.702240  &  1.224899  &  16.201 $\pm$  0.002  &   1.801  $\pm$  0.006  &    2.206  $\pm$  0.007 &   12.172  $\pm$  0.018  &  11.303  $\pm$  0.025  &  11.019  $\pm$   0.025  &  10.808  $\pm$  0.002  &         $...$   $~~~~~~$      \\
   14  &  101.660364  &  1.219329  &  15.163 $\pm$  0.002  &   0.658  $\pm$  0.005  &    0.897  $\pm$  0.013 &   13.664  $\pm$  0.023  &  13.438  $\pm$  0.036  &  13.302  $\pm$   0.042  &  13.218  $\pm$  0.009  &         $...$    $~~~~~~$     \\
   15  &  101.695928  &  1.240098  &  11.657 $\pm$  0.001  &   0.331  $\pm$  0.001  &    0.428  $\pm$  0.001 &   11.001  $\pm$  0.018  &  10.895  $\pm$  0.025  &  10.821  $\pm$   0.021  &  10.863  $\pm$  0.002  &         $...$    $~~~~~~$     \\
   16  &  101.697027  &  1.251858  &  16.037 $\pm$  0.043  &   1.509  $\pm$  0.055  &    1.611  $\pm$  0.068 &   12.930  $\pm$  0.026  &  12.423  $\pm$  0.034  &  12.075  $\pm$   0.034  &  12.083  $\pm$  0.004  &         $...$   $~~~~~~$      \\
   17  &  101.707273  &  1.250715  &  18.082 $\pm$  0.005  &   2.273  $\pm$  0.028  &    2.860  $\pm$  0.006 &   12.790  $\pm$  0.018  &  11.691  $\pm$  0.026  &  11.269  $\pm$   0.025  &  10.937  $\pm$  0.003  &         $...$   $~~~~~~$      \\
   18  &  101.781223  &  1.227141  &  15.286 $\pm$  0.004  &   1.804  $\pm$  0.009  &    2.248  $\pm$  0.003 &   11.254  $\pm$  0.019  &  10.401  $\pm$  0.020  &  10.126  $\pm$   0.019  &   9.938  $\pm$  0.002  &         $...$   $~~~~~~$      \\
   19  &  101.756836  &  1.248329  &  15.473 $\pm$  0.003  &   0.840  $\pm$  0.003  &    1.099  $\pm$  0.003 &   13.624  $\pm$  0.019  &  13.266  $\pm$  0.024  &  13.119  $\pm$   0.030  &  13.093  $\pm$  0.008  &         $...$   $~~~~~~$      \\
   20  &  101.755078  &  1.237696  &  14.601 $\pm$  0.002  &   0.952  $\pm$  0.004  &    1.178  $\pm$  0.005 &   12.565  $\pm$  0.021  &  12.172  $\pm$  0.020  &  12.057  $\pm$   0.021  &  12.030  $\pm$  0.005  &         $...$    $~~~~~~$     \\
   21  &  101.783209  &  1.238542  &  14.248 $\pm$  0.003  &   0.483  $\pm$  0.002  &    0.690  $\pm$  0.003 &   13.137  $\pm$  0.023  &  12.998  $\pm$  0.030  &  12.899  $\pm$   0.026  &  12.834  $\pm$  0.007  &         $...$  $~~~~~~$       \\
   22  &  101.768600  &  1.242802  &  16.683 $\pm$  0.003  &   1.446  $\pm$  0.006  &    2.174  $\pm$  0.004 &   13.213  $\pm$  0.021  &  12.608  $\pm$  0.020  &  12.386  $\pm$   0.024  &  12.140  $\pm$  0.005  &         $...$   $~~~~~~$      \\
   23  &  101.768194  &  1.248966  &  16.017 $\pm$  0.004  &   1.770  $\pm$  0.005  &    2.305  $\pm$  0.002 &   11.783  $\pm$  0.019  &  10.860  $\pm$  0.020  &  10.555  $\pm$   0.021  &  10.323  $\pm$  0.002  &         $...$   $~~~~~~$      \\
   24  &  101.792152  &  1.249625  &  13.821 $\pm$  0.004  &   0.549  $\pm$  0.002  &    0.740  $\pm$  0.003 &   12.674  $\pm$  0.019  &  12.402  $\pm$  0.020  &  12.344  $\pm$   0.023  &  12.299  $\pm$  0.005  &         $...$    $~~~~~~$     \\
   25  &  101.783535  &  1.271522  &  15.903 $\pm$  0.003  &   1.693  $\pm$  0.005  &    2.090  $\pm$  0.003 &   12.141  $\pm$  0.019  &  11.370  $\pm$  0.020  &  11.099  $\pm$   0.021  &  10.916  $\pm$  0.003  &  10.907  $\pm$  0.003\\
   26  &  101.810802  &  1.277135  &  15.134 $\pm$  0.002  &   0.739  $\pm$  0.005  &    0.936  $\pm$  0.005 &   13.607  $\pm$  0.025  &  13.327  $\pm$  0.038  &  13.150  $\pm$   0.034  &  13.165  $\pm$  0.008  &  13.132  $\pm$  0.010\\
   27  &  101.813572  &  1.282032  &  18.224 $\pm$  0.027  &   1.899  $\pm$  0.028  &    2.541  $\pm$  0.008 &   13.647  $\pm$  0.026  &  12.714  $\pm$  0.026  &  12.393  $\pm$   0.024  &  12.085  $\pm$  0.004  &  12.061  $\pm$  0.005\\
   28  &  101.750326  &  1.247554  &  14.367 $\pm$  0.004  &   0.709  $\pm$  0.003  &    0.858  $\pm$  0.006 &   12.931  $\pm$  0.025  &  12.598  $\pm$  0.028  &  12.536  $\pm$   0.027  &  12.502  $\pm$  0.006  &         $...$   $~~~~~~$      \\
   29  &  101.749411  &  1.261315  &  13.565 $\pm$  0.005  &   1.148  $\pm$  0.003  &    1.307  $\pm$  0.004 &   11.442  $\pm$  0.021  &  10.831  $\pm$  0.020  &  10.737  $\pm$   0.023  &  10.687  $\pm$  0.002  &         $...$    $~~~~~~$     \\
   30  &  101.772527  &  1.267760  &  15.406 $\pm$  0.004  &   0.834  $\pm$  0.003  &    1.081  $\pm$  0.003 &   13.596  $\pm$  0.023  &  13.155  $\pm$  0.028  &  13.017  $\pm$   0.037  &  13.001  $\pm$  0.008  &  12.995  $\pm$  0.010\\
   31  &  101.740814  &  1.270879  &  17.872 $\pm$  0.005  &   2.871  $\pm$  0.041  &    3.932  $\pm$  0.003 &   10.489  $\pm$  0.019  &   8.853  $\pm$  0.039  &   8.118  $\pm$   0.015  &   7.623  $\pm$  0.029  &   7.629  $\pm$  0.011\\
   32  &  101.783952  &  1.276780  &  17.428 $\pm$  0.008  &   2.170  $\pm$  0.016  &    2.688  $\pm$  0.003 &   12.529  $\pm$  0.023  &  11.460  $\pm$  0.028  &  11.071  $\pm$   0.023  &  10.818  $\pm$  0.002  &  10.815  $\pm$  0.003\\
   33  &  101.790469  &  1.288697  &  12.782 $\pm$  0.002  &   0.597  $\pm$  0.003  &    0.724  $\pm$  0.003 &   11.595  $\pm$  0.019  &  11.262  $\pm$  0.020  &  11.226  $\pm$   0.017  &  11.214  $\pm$  0.003  &  11.210  $\pm$  0.004\\
   34  &  101.797372  &  1.315962  &  12.522 $\pm$  0.002  &   0.774  $\pm$  0.002  &    0.918  $\pm$  0.002 &   11.018  $\pm$  0.019  &  10.597  $\pm$  0.020  &  10.511  $\pm$   0.021  &  10.501  $\pm$  0.002  &  10.472  $\pm$  0.003\\
   35  &  101.763612  &  1.311435  &  13.608 $\pm$  0.004  &   0.417  $\pm$  0.005  &    0.765  $\pm$  0.004 &   12.270  $\pm$  0.018  &  12.075  $\pm$  0.020  &  11.984  $\pm$   0.021  &  11.868  $\pm$  0.004  &  11.845  $\pm$  0.005\\
\hline\end{tabular}
\end{table*}

\subsubsection{Slitless Spectroscopy}

The grism slitless spectroscopic observations were obtained to identify $H_\alpha$ emission line stars using HFOSC of HCT on 2011 January 25 and 2013 November 08. The observations were carried out using a combination of  Grism 5 (5200-10300 {\AA}) or Grism 8 (5800-8350 {\AA}) and the $H_\alpha$ broadband filter (6300-6740{\AA}). The log of observations is shown in Table~\ref{tab:observation}. The $H_\alpha$ emitting sources were  identified by the presence of bright spot along the slitless spectra, while non-emitting $H_\alpha$ sources do not show any bright spot. Thus, we  have identified  16 sources as   $H_\alpha$ emission line sources  which are discussed later in sect. 3.5.3.

\subsection{ARCHIVAL DATA SETS}

\subsubsection{Near-Infrared Data form UKDISS and 2MASS}  
Near-IR $JHK$ photometric data towards NGC 2282 were acquired from the UKIRT Infrared Deep Sky Survey
(UKIDSS, Lawrence et al. 2007), which were taken during the UKIDSS Galactic Plane Survey (GPS)
(Lucas et al. 2008; data release 6). The UKIDSS GPS has saturation limits at $J$ = 13.25, $H$ = 12.75 and $K$ = 12.0 mag, respectively (Lucas et al. 2008). We therefore, replaced the  saturated stars  with the  2MASS PSC data. We set the 2MASS limit 0.5 magnitudes fainter than UKIDSS saturation limits following Alexander et al. (2013). In total, 150 saturated sources in UKIDSS catalog were replaced by 2MASS sources.

\begin{table*}
\caption{Details of the spectroscopically studied stars.}
\begin{threeparttable}

\label{tab:spec}
\begin{tabular}{cccccccccccc}

\hline 
\multicolumn{1}{c}{ID} & \multicolumn{1}{c}{RA (J2000)} & \multicolumn{1}{c}{Dec (J2000)} & \multicolumn{1}{c}{$V$} &  \multicolumn{1}{c}{$V-I$} & \multicolumn{1}{c}{Spectroscopic} & \multicolumn{1}{c}{Photometric}\tnote{*} & \multicolumn{1}{c}{distance}\tnote{**} & \multicolumn{1}{c}{distance}& \multicolumn{1}{c}{Spectral} & \multicolumn{1}{c}{Remarks} \\ 
\multicolumn{1}{c}{} & \multicolumn{1}{c}{(h:m:s)} & \multicolumn{1}{c}{(d:m:s)} & \multicolumn{1}{c}{(mag)} &  \multicolumn{1}{c}{(mag)} & \multicolumn{1}{c}{$A_V$ (mag)} & \multicolumn{1}{c}{$A_V$ (mag)} & \multicolumn{1}{c}{modulus} & \multicolumn{1}{c}{(pc)} & \multicolumn{1}{c}{Type}& \multicolumn{1}{c}{} \\ \hline
\hline
1 & 06:46:50.304 & +1:16:35.904 & 14.674 & 1.438 & 4.91 $\pm$ 0.20 & 4.69 $\pm$ 0.90 & 11.01 $\pm$ 0.20 & 1592 $\pm$ 147 &B5V&member\\
2 & 06:46:57.282 & +1:17:45.491 & 13.858 & 0.945 & 0.64 $\pm$ 0.15 & $...$  & 7.96 $\pm$ 0.15 & 390 $\pm$ 30 &G8V & foreground\\
3 & 06:46:50.544 & +1:18:04.859 & 14.285 & 1.177 & 0.67 $\pm$ 0.20 & $...$ & 7.71 $\pm$ 0.20 & 348 $\pm$ 32 & K1V &foreground\\
4 & 06:46:40.092 & +1:18:47.447 & 13.348 & 0.735 & 0.46 $\pm$ 0.16 & $...$  &9.20 $\pm$ 0.17 &692 $\pm$ 50 &F7V &foreground\\
5 & 06:46:45.986 & +1:18:07.775 & 13.777 & 0.803 & 0.59 $\pm$ 0.17 & $...$ & 9.21 $\pm$ 0.18 & 695 $\pm$ 55 &F9V  &foreground\\
6 & 06:46:39.456 & +1:18:38.500 & 14.037 & 0.809 & 0.86 $\pm$ 0.21 & $...$ & 10.33 $\pm$ 0.21 & 1165 $\pm$ 110 & F0V &foreground\\
7 & 06:46:51.581 & +1:18:57.756 & 10.198 & 0.337 & 1.66 $\pm$ 0.17 & 1.61 $\pm$ 0.71 & 11.18 $\pm$ 0.18 & 1722 $\pm$ 135 & B2V & HD 289120; member\\
8 & 06:46:56.424 & +1:16:40.944 & 15.934 & 1.841& 8.82 $\pm$ 0.14 & 10.72 $\pm$ 0.62 &$...$&$...$ & B0.5Ve & Herbig Be; member\\
\hline\end{tabular}
\begin{tablenotes}\footnotesize
 \item [*] Photometric $A_V$'s are calculated from extinction map. Foreground stars have not considered here.
  \item [**] Distance moduli refers to the intrinsic distance moduli obtained from near-IR apparent distance moduli.
\end{tablenotes}
\end{threeparttable}
\end{table*}

\subsubsection{{\it Spitzer}-IRAC data from warm mission}

The IRAC observations in 3.6 and 4.5 $\mu$m bands (channels 1 and 2) were available in the {\it Spitzer} archive program (Program ID: 61071; PI: Whitney, Barbara A). The data sets were taken towards NGC 2282  on May 25, 2011  at  various dithered positions and with integration time of 0.4 and 10.4 sec per dither. The  basic calibrated Data (version S18.18.0) were downloaded from Spitzer archive\footnote{http://archive.spitzer.caltech.edu/}. The raw data were processed and the final mosaic frames were created using MOPEX (version 18.5.0) with an image scale of 1.2 arcsec pixel$^{-1}$. We performed point response function (PRF) fitting method using  APEX tool provided by {\it Spitzer} Science  centre on all the {\it Spitzer} IRAC images to extract the magnitudes of point sources. The detailed procedure of source detection and magnitude extraction is described in Jose et al. (2013). We adopted zero  point flux densities  of 280.9 and 179.7 Jy for the 3.6 and 4.5 $\mu$m bands, respectively, following the Warm Spitzer Observer Manual. We finally detected 3049 and 2341  number of sources within 7$\arcmin$ radius around the cluster in IRAC 3.6 and 4.5 $\mu$m bands, respectively. The IRAC data of two bands were matched with a radial matching tolerance of 1.2 arcsec. Thus our final IRAC catalog contains 3304 sources, of  which, 2085 sources are detected in both bands.

\subsubsection{IPHAS data} 
INT Photometric $H_\alpha$ Survey of the Northern Galactic Plane (IPHAS) is a photometry survey using wide field camera (WFC) on the 2.5m Isaac Newton Telescope (INT) with Sloan $r$, $i$ filters and $H_\alpha$ narrow-band filter  (Drew et al. 2005; Gonzalez-Solares et al. 2008). The data for NGC 2282 were obtained in all three bands from  the data release 2 (Barentsen et al. 2014).

 \subsubsection{Multiwavelength Catalog} 
 
The multiwavelength catalog was built by cross-matching all the catalogs described in the above sections  except IPHAS data. The $JHK$ and IRAC  data were matched with a matching radius of  2$\arcsec$. Before this, we performed several test matches by increasing the matching  radius in step of 0.1$\arcsec$  from 1.0$\arcsec$ to 3.5$\arcsec$. We found that the matching radius of 2$\arcsec$ is sufficient for cross-matching two catalogs. In few cases, we got more than one matching sources, and we have taken  the closest one as the best match. Following the same method, we adopted the matching radius of  2$\arcsec$ to match between IR and optical catalog. Finally, the results of each match have been visually inspected in Optical, UKIDSS, 2MASS and {\it Spitzer} images. Our final catalog contains total 5601  number of sources within 7$\arcmin$ radius around the cluster, but all of them do not have  detection in all  wavelengths. The entire photometric catalog is presented in Table~\ref{tab:catalog}. But, those sources with uncertainty $\le$ 0.1 mag have been taken for our study to ensure good photometric accuracy.

 The completeness limits at various bands were estimated from histogram turn over method (e.g. Samal et al. 2015). We considered $\sim$ 90\% completeness of our data from the turning points of magnitudes at which cumulative logarithmic distribution of sources in the histograms deviate from linear distribution (figures are not shown). We found that the photometric data is complete down to $V$ = 21 mag, $I$ = 20 mag, $J$ = 18.5 mag, $H$ = 18 mag, $K$ = 17.5 mag, [3.6] = 15.5 mag and [4.5] = 15 mag, respectively. However, completeness is limited by various factors such as, variable reddening, central luminous sources, stellar crowding across the region etc. Bright extended sources, variable nebulosity, significant saturation in {\it Spitzer}-IRAC bands also limits the point source detections. The modest sensitivity of UKIDSS and {\it Spitzer}-IRAC observations significantly limits our study.

      For cross-checking, the completeness limits of $V$-band was also  estimated by the method of inserting artificial stars of various magnitude bins into the image using IRAF (e.g. Jose et al. 2013). The frames were reduced using the same procedure used for original frame. The $V$-band photometry was 100\% complete down to 18 mag,  reduced to $\sim$ 90\% for 19-20 mag range, and 78\% for the 21-21.5 mag range. The completeness limit obtained for $V$-band using artificial star injection method fairly matches with that of the histogram analysis.

\section{Analysis and Results}

\subsection{Stellar density and cluster radius }

\begin{figure}

\includegraphics[width=8.0 cm,height=8.0cm, angle=0]{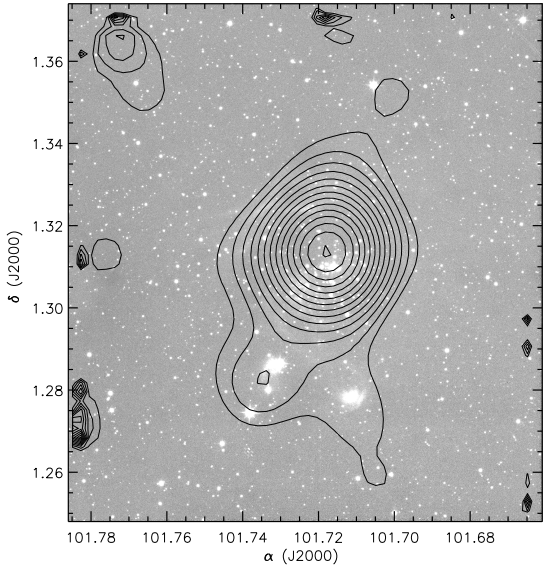}
  \caption{Stellar surface density map of sources detected in K-band towards NGC 2282 region overlaid on the UKIDSS $K$-band image. Stellar surface density was calculated using the nine nearest neighbors. }  
  \label{fig:st_den}  
\end{figure}

In order to understand the spatial structure of the cluster, we generated the stellar surface density map using the nearest neighborhood technique and following the method introduced by Casertano \& Hut (1985). Briefly, the stellar density $\sigma(i, j)$ inside a cell of an uniform grid with centre at the coordinates (i, j) is
$\sigma(i, j)$ = $N-1 \over \pi r^2_N(i, j)$
where, $r_N$ is the distance from the centre of the cell to the
N$^{th}$ nearest source. Value of N is allowed to vary depending upon how smallest scale structures of the field are interested to study. In Fig.~\ref{fig:st_den}. we have shown the surface density map which is estimated using the $K$- band  data ($K$ $\leq$ 17.5 mag with photometric uncertainty of $<$ 0.1) for the stars towards NGC 2282 region overlaid on the UKIDSS $K$-band image. This map has been obtained using a grid size of 10$\arcsec$ $\times$ 10$\arcsec$ and N=9. As evident in the IR image of the cluster (see Fig.~ \ref{fig:optical_nir}b),  the surface density map shows a more or less centrally concentrated clustering  with a slight elongation along the north-south direction. We obtained the peak stellar density as  418 $stars/pc^2$ (96 $stars/arcmin^2$). The radius of the cluster can be considered as the semi-major axis of the outermost elliptical contour in Fig.~\ref{fig:st_den}. The cluster radius was thus estimated as $\sim$ 3.15$\arcmin$, which agrees well with the literature value (Horner et al. 1997).

\subsection{Spectral Classification of Optically Bright Sources}

In this section we estimate the   spectral types of  8 bright objects  observed towards NGC 2282. The targets for low-resolution spectroscopy were selected on the basis of their brightness ($J$ $<$ 13 mag) around the  cluster (see Fig.~\ref{fig:optical_nir}). The coordinates and optical magnitudes of these sources are given in Table~\ref{tab:spec}, and the  flux calibrated,  normalized spectra are shown in Fig.~\ref{fig:spec1}.

\begin{figure*}
\includegraphics[width=15.0 cm,height=10.0cm]{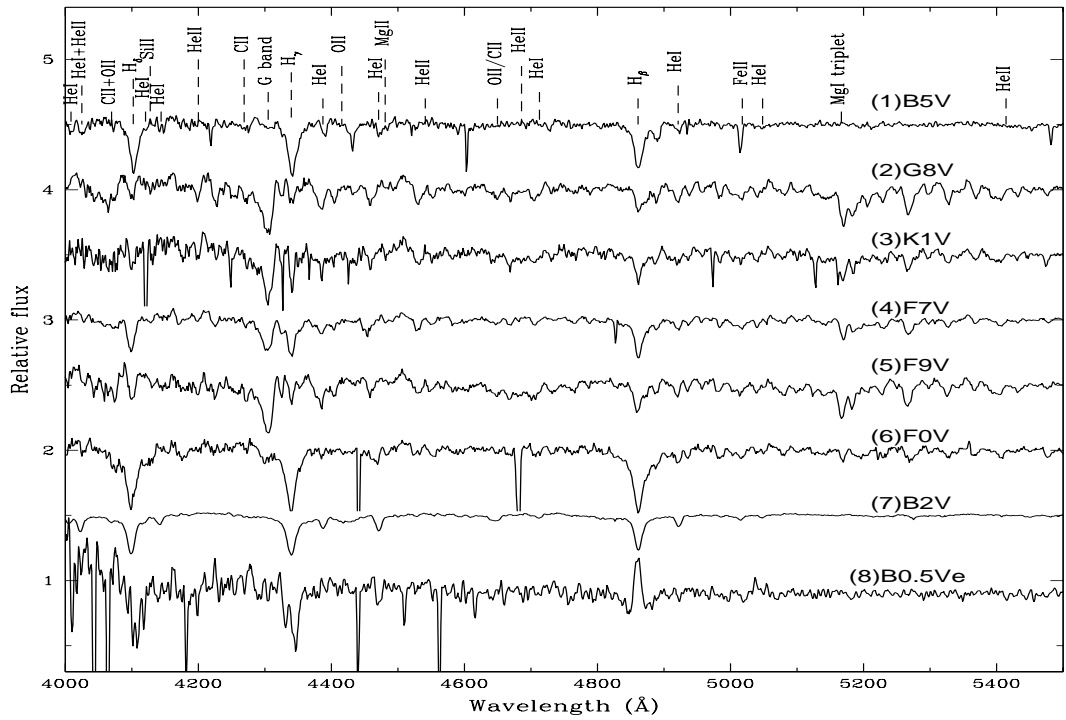}

  \caption{Flux-calibrated normalized spectra for the optically bright sources in NGC 2282. Important spectral lines are marked. The star IDs and spectral classes identified are given in the figure.}
   \label{fig:spec1}
\end{figure*}

\addtocounter{figure}{-1}
\begin{figure*}
\includegraphics[width=15.0 cm,height=10.0cm]{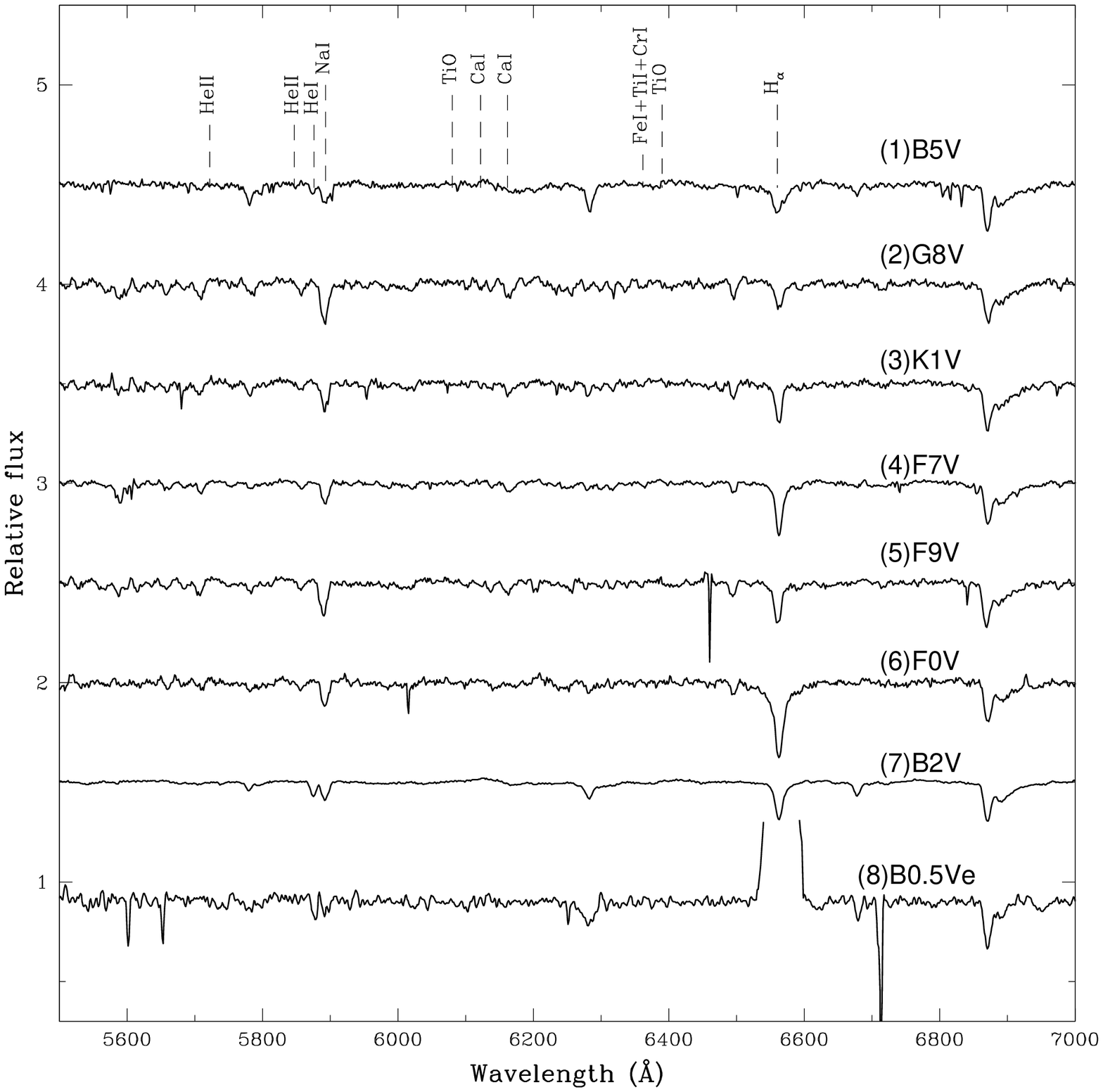}
  \caption{continued...(H$_\alpha$ emission line strength is not shown for ID = 8 as the  intensity is very high.)}
  \label{fig:spec1}
\end{figure*}

We classified the observed spectra using different spectral libraries available in the literature (Walborn \& Fitzpatrick 1990; Jacoby, Hunter \& Christian 1984; Torres-Dodgen \& Weaver 1993; Allen \& Strom 1995). First, we determined a specific spectral range from strong conspicuous features. For  e.g., the  absence of \heii 5411 \AA, in a spectra constraints the spectral type as B0.5 or later. While the absence of  \hei 5876 \AA, in a spectra limits the spectral type to A0 or later (Lundquist et al. 2014). Spectra of O- and B-type stars have the features of hydrogen and helium along with other atomic lines (e.g. \oii, \ciii, \siiii, \siiv, \mgii). The \heii line strength appears weaker for late O-type stars and \heii 4686 is last visible in B0.5  (Walborn \& Fitzpatrick 1990). If a spectrum shows \heii line at 4200 \AA$~$ along with the \oii/\ciii blend at 4650 \AA, we can classify them to be earlier than B1. The absence of \heii 4200, \heii 4686, \mgii 4481 and presence of weak features of silicon along with the weak \oii/\ciii blends at 4070 and 4650 \AA, indicates the spectral type in the range of B1$-$B2. For B2-type spectra, \hei is in its maximum strength, but for later-type stars \siii 4128-4130 and \mgii gets stronger (Walborn \& Fitzpatrick 1990). The presence of \hei lines indicates that spectral type is earlier than B5$-$B7. The late-type sources are classified using the spectral lines TiO 5847-6058, \nai 5893, \hei, $H_{\alpha}$, \cai 6122, 6162, \feii 6456. A- and F- stars are identified comparing their $H_\alpha$ equivalent width. The G-band (CH $\lambda$4300 \AA) appears from F-type stars. G-type stars are classified in comparisons with the equivalent width of $H_{\alpha}$ and \mgi triplet ($\lambda \lambda$ 5167, 5172, 5183 \AA). Finally, each source was compared visually to the standard library spectra from Jacoby et al. (1984).  However, on the basis of the low$-$resolution spectroscopy of early type stars, it is difficult to distinguish the luminosity class  between supergiants, giants, dwarfs and pre$-$main sequence stars. 

Comparison with the standard spectral libraries of Jacoby et al. (1984) and Walborn \& Fitzpatrick (1990), the star ID 1 is classified as B5 V as it has no features of \heii 4200,  weak features of \heii 4541 and C II 4267.
The star IDs 2, 3, 4, 5 and 6 are classified as G8 V, K1 V, F7 V, F9 V and F0 V, respectively, as these stars show presence of G-band (CH 4300 \AA) and \mgiii triplet ($\lambda\lambda$ 5167 \AA, 5177 \AA, 5183 \AA), and their luminosity classes resemble better with main-sequence stars rather than super-giants or giants. The star ID 7 is classified as B2 V as it shows weak features of \siiii 4552 along with blended \ciii/\oii 4070 and \ciii/\oii 4650, and  absence of \heii 4200, \heii 4686, and \mgii 4481. The star ID 8 is classified as B0.5 Ve as it shows ionized helium lines at  \heii 4200, \heii 4686, \heii 5411,  and \heii 5720 including strong H$_\alpha$, H$\beta$ and H$\gamma$ lines in emission. Based on low-resolution spectra of our targets, an uncertainty of $\pm$1 or more in the sub$-$class estimation  is expected. Photometric and spectroscopic details  of all the  8 sources are  given in Table~\ref{tab:spec}.

\subsection{Reddening towards the Region}

\begin{figure}
\includegraphics[width=8.0 cm,height=8.0 cm]{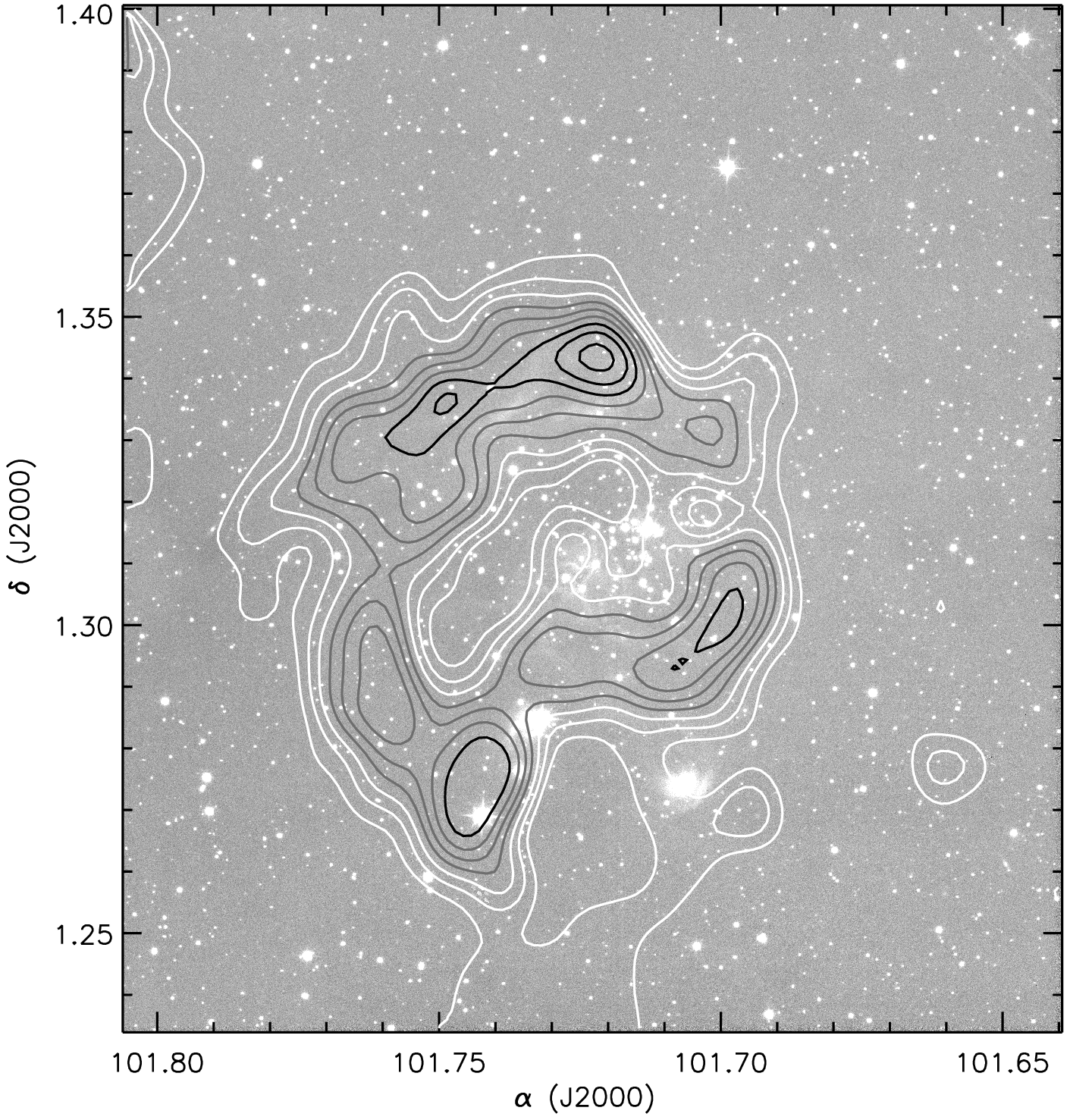}
  \caption{Extinction map over plotted on the UKIDSS $K$-band image. $A_K$ values are estimated from the ($H-K$) colours. The contour levels are for  $A_K$ values 0.32 - 0.42 (white), 0.47 - 0.57 (grey) and 0.62 - 0.82 mag (black), respectively.}
  \label{fig:ext_map}
\end{figure}

 The extinction in an embedded cluster is distributed non-uniformly. It is important to know the spatial variation of extinction of the cluster to characterize the cluster members.  We estimated the $K$-band extinction towards NGC~2282 using 2MASS and UKIDSS data  within an area  $\sim$ 10$\arcmin \times$ 10$\arcmin$, to understand the local extinction towards the region. We measured the $A_K$ value using ($H-K$) colours of the stars. The sources without infrared excess ( i.e., background dwarfs and non-excess sources within the field)  were used to generate the extinction map. Following Gutermuth et al. (2005), we used the grid method to measure the mean value of $A_K$ (see Jose et al. 2013 for details). Briefly, we divided the region of our interest into small grids of size 10$\arcsec$ $\times$ 10$\arcsec$. The mean value and standard deviation of $(H-K)$ colours of 5 nearest neighbor stars from the centre of each grid was measured. We rejected any sources deviating above 3$\sigma$ to calculate the mean value of ($H-K$) at each grid position. To eliminate foreground extinction, we took only those stars having $A_K$ $>$ 0.12 mag to generate the extinction map (Jose et al. 2013). The ($H-K$) values were converted into $A_K$ using the reddening law given by Falherty et al. (2007),  i.e.,  $A_K = 1.82 \times (H-K)_{obs} - (H-K)_0$, where $(H-K)_0$ is the average intrinsic colour of stars, which is assumed to be 0.2  (Allen et al. 2008; Gutermuth et al. 2009). To improve the quality of the extinction map, we excluded the probable YSOs candidates (see sect. 3.5), which otherwise might show high extinction value due to  near-IR excess from circumstellar disc emission. The derived extinction map is shown in Fig.~\ref{fig:ext_map}. The  extinction within NGC 2282 varies between $A_V$ = 1.6$ - $8.7 mag with an average extinction of $A_K$ $\sim$ 0.35, which corresponds to $A_V$ $\sim$ 3.9 mag considering the extinction ratio $A_K$/$A_V$ = 0.090 given by Cohen et al. (1981). The south-eastern part of the cluster found to be  at relatively high extinction ($A_K$  is $\sim$ 0.5 mag) compared to the  average extinction of the cluster.

We also verified the extinction to each  spectroscopically observed source from their photometry. According to the spectral types  given in Table~\ref{tab:spec}, we estimated the average interstellar extinction towards each star. We first calculated the IR colour excesses $E(J-H)$, $E(H-K)$, $E(J-K)$ for each sources using their observed colours. We transformed these colour excess into visual extinction according to the  extinction law, $A_V$ = $E(J-H)/0.11$, $A_V$ = $E(H-K)/0.065$, $A_V$ = $E(J-K)/0.175$ (Cohen et al. 1981). Finally, we took the average   $A_V$ values for each star and are given in Table~\ref{tab:spec}. The extinction of HD 289120, which is the main illuminating stars of the cluster, is $A_V$ $\sim$ 1.65 mag.

\subsection{Distance and Membership of the Bright Sources}

The projected stars against NGC 2282 could be either the young members of the clusters, background stars or foreground stars. We estimated the membership of the candidates based on their spectral types, distance, and photometry. One can derive the distance to a star from estimated spectral type, apparent magnitudes and extinction. We calculated the optical  intrinsic distance modulus $(V_0-M_V)$ and also the near-IR intrinsic distance modulus  $(J_0-M_J)$, $(H_0-M_H)$ and $(K_0-M_K)$. We prefer near-IR intrinsic distance modulus over optical as it is relatively less uncertain on extinction and come from simultaneous three band measurements of 2MASS data. From near-IR intrinsic distance modulus, we estimated the distance to each spectroscopically observed stars and are given in Table~\ref{tab:spec}. The intrinsic colours and absolute magnitudes are taken from Koorneef (1983), Schmidt-Kaler (1982) and Pecaut \& Mamajek et al. (2013). The values of intrinsic distance modulus for star IDs 1 and 7 are found to be 11.01 $\pm$ 0.20 mag, 11.18 $\pm$ 0.18 mag,                                                                                                                                                                                                                                                                                                                                                                                                                                                                                                                                                                                                                                                                                                                                                                                                                                                                                                                                                                                                                                                                                                                                                                                                                                                                                                                                                                                                                                                                                                                                                                                                                                                                                                                                                                                                                                                                                                                                                                                                                                                                                                                                                                                                                                                                                                                                                                                                                                                     which corresponds to 1592 $\pm$ 147 pc and 1722 $\pm$ 135 pc, respectively. Three early B-type stars are adopted as members of the cluster. Thus, we measure the average distance to the cluster as $\sim$ 1650 $\pm$ 100 pc from our spectrophotometry observations. Our estimated distance agrees well with the published value in the literature (Racine 1968; Avedisova \& Kondratenko 1984; Horner et al. 1997).  Considering their distances, the star IDs 2, 3, 4, 5 and 6 seem to be foreground stars. The distance to the star ID 8 could not be measured accurately, as it is an emission line star with large infrared excess due to the presence of circumstellar disc.

\begin{figure}
\includegraphics[width=8.0 cm,height=8.0 cm]{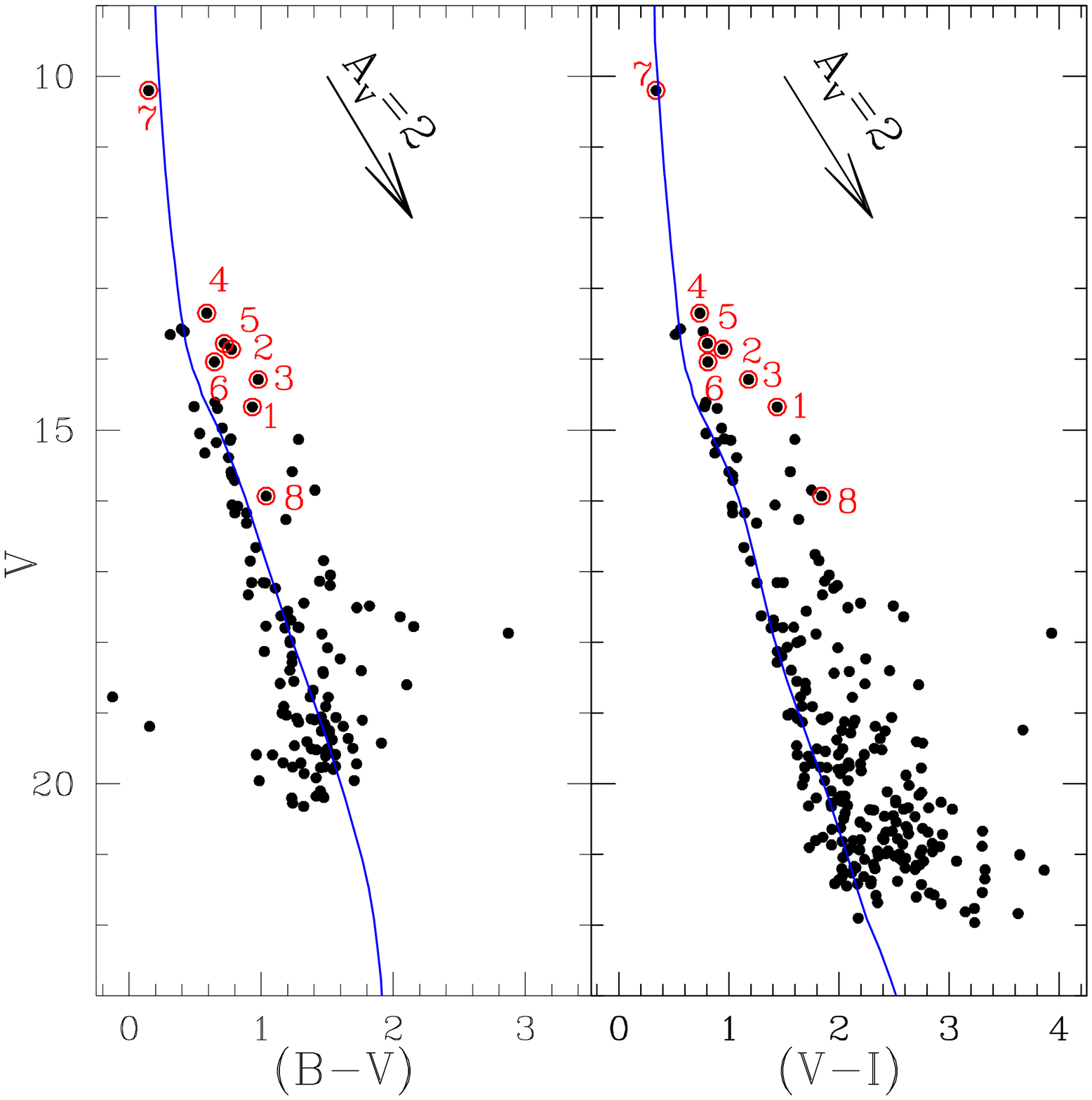}
  \caption{Optical CMDs for all the stars  within 3.15$\arcmin$ radius of the cluster. The solid blue curve is the ZAMS from Girardi et al. (2002), corrected for the distance of 1.65 kpc and reddening $E(B-V)$ = 0.52 mag. The reddening vector $A_V$ = 2 mag is also shown. The sources marked with red circles are the spectroscopically studied stars towards NGC 2282.}
  \label{fig:spec_viv}
\end{figure}

 In Fig.~\ref{fig:spec_viv}, we show $V/(B-V)$ and $V/(V-I)$ colour-magnitude diagrams (CMDs) for all the stars detected in optical photometry towards NGC~2282. Spectroscopically classified stars are also marked with red circles. The solid blue curve is the zero-age main sequence (ZAMS) by Girardi et al. (2002) shifted for the distance 1.65 kpc and reddening $E(B-V)$ = 0.52 mag and $E(V-I)$ = 0.65 mag ($E(B-V)/E(V-I$) = 1.25 ; Cohen et al. 1981), respectively. Since the cluster reddening is highly variable, we used the extinction of the main-sequence (MS) member, HD 289120, located at the cluster centre for corrections of theoretical isochrones. HD 289120 lie on the ZAMS locus. The star IDs 1 and 8 are also high mass members of the cluster. Though the  stars with IDs 2, 3, 4, 5 and 6 fall towards right side of the ZAMS, spectroscopic analysis  reveals that these stars could be foreground towards NGC 2282. For more reliable membership analysis, we used various observable signature of youthfulness such as emission at $H_\alpha$, excess emission due to presence of disc to identify the probable members of the cluster (see sect. 3.5).

\subsection{Identification and Classification of YSOs}

We identify and classify the YSOs towards NGC 2282 based on their IR colours. Following Gutermuth et al. (2008, 2009), we used IRAC 3.6 and 4.5 $\mu$m band data along with $H$ and $K$ data to identify and classify the YSOs. The main limitations for the identification of YSOs based on the IR colours arise from the contaminations of different non-stellar sources in IR detections such as, extragalactic objects like polycyclic aromatic hydrocarbon (PAH) emitting and star-forming galaxies, active galactic nuclei (AGN), unresolved knots of shock emission, PAH-emission contaminated apertures, etc. These sources have considerable IR colours, which could mimic the colours of YSOs in colour-colour (CC) diagram. A number of candidate YSOs could be missed in the mid-IR bands due to the limited sensitivity of IRAC observations. Hence, we identify more YSOs using their near-IR colours (UKIDSS and 2MASS) from the master catalog given in Table~\ref{tab:catalog}. We used $(J-H)/(H-K)$ near-IR CC diagram  to identify the additional YSOs, but we cannot classify them in to  Class I or Class II category based on their  near-IR colours alone. Similarly, the presence of H$_\alpha$ emission  is considered as a significant characteristic of a YSO with ongoing disc accretion process (e.g. Dahm 2005). We identified the H$_\alpha$ emission line stars from our slitless spectroscopy data and IPHAS photometry survey  (see sect. 3.5.3.). Below we explain the details about the various YSO selection processes.

\begin{figure}
\includegraphics[width=8.0 cm,height=8.0 cm]{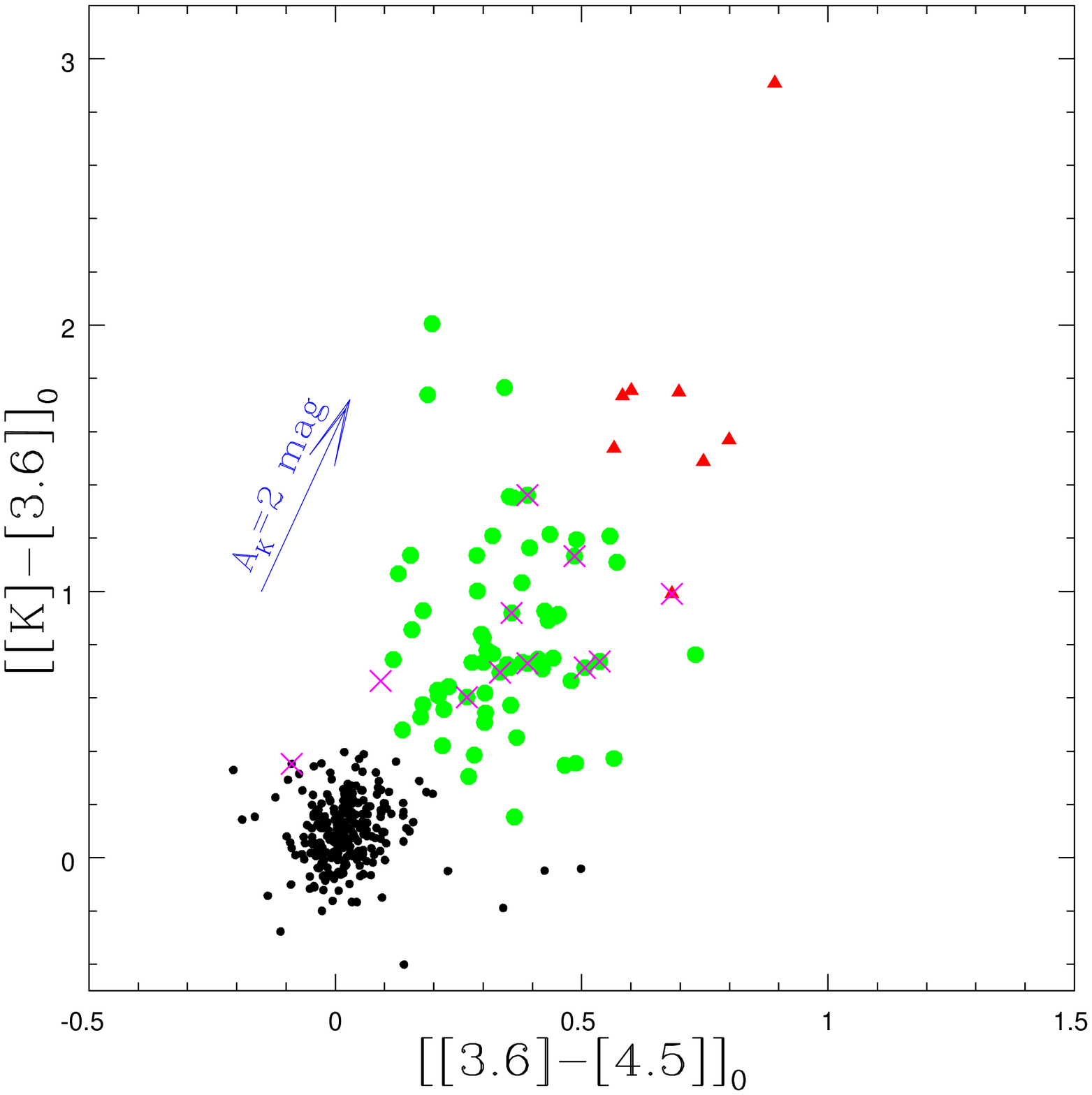}
  \caption{Dereddened $[K-3.6]_0$ vs $[3.6-4.5]_0$ CC diagram for all the sources within the  cluster radius  of 3.15$\arcmin$ after removing contaminants. The reddening vector $A_K$ = 2 mag is plotted by using the reddening law from Flaherty et al. 2007. The sources in green and red are candidate Class II and Class I sources, respectively. The $H_\alpha$ emission sources detected from slitless spectroscopy are marked as magenta crosses.}
    \label{fig:irac}
\end{figure}

\begin{figure*}
\includegraphics[width=8.0 cm,height=8.0 cm]{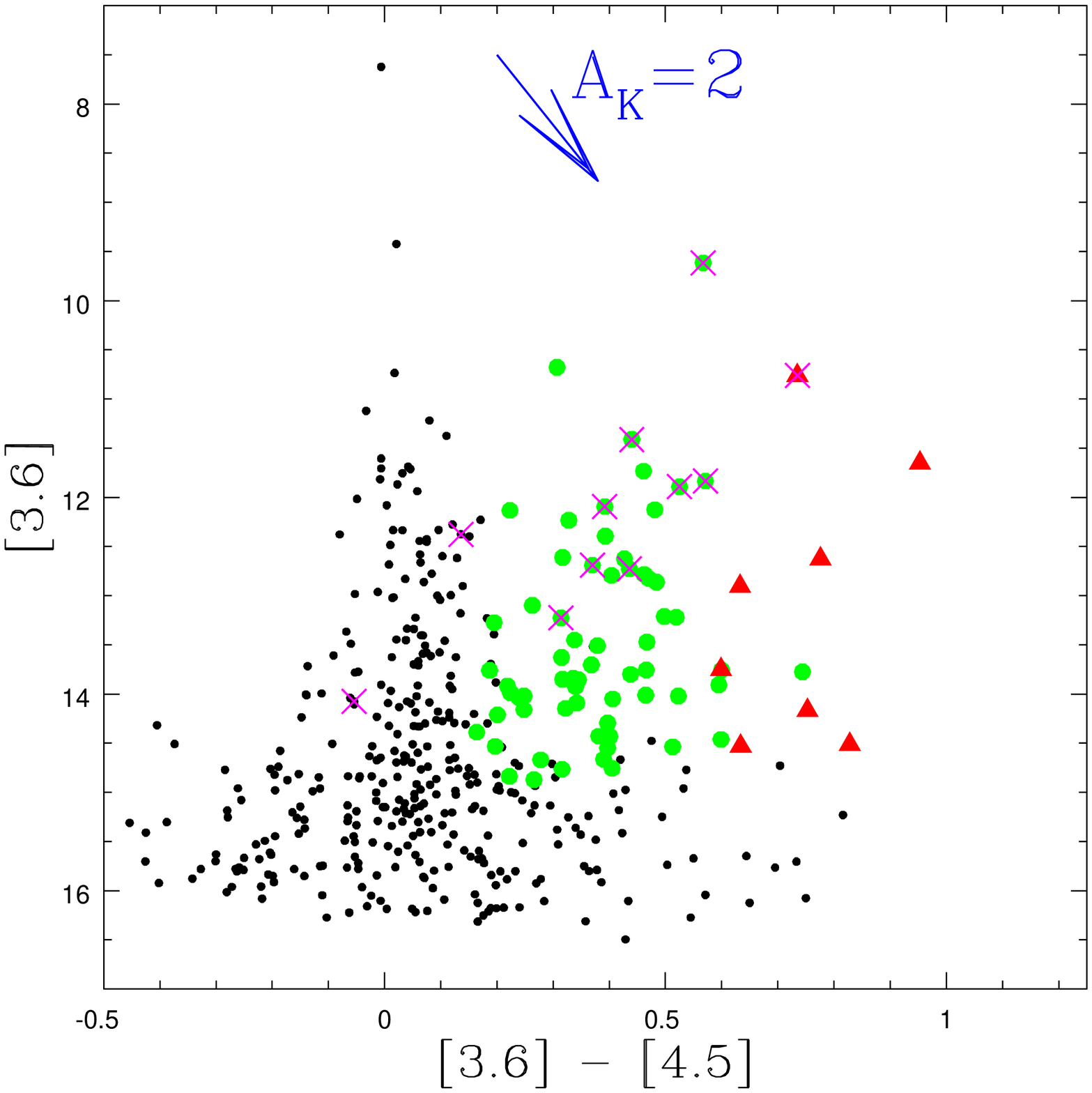}
\includegraphics[width=8.0 cm,height=8.0 cm]{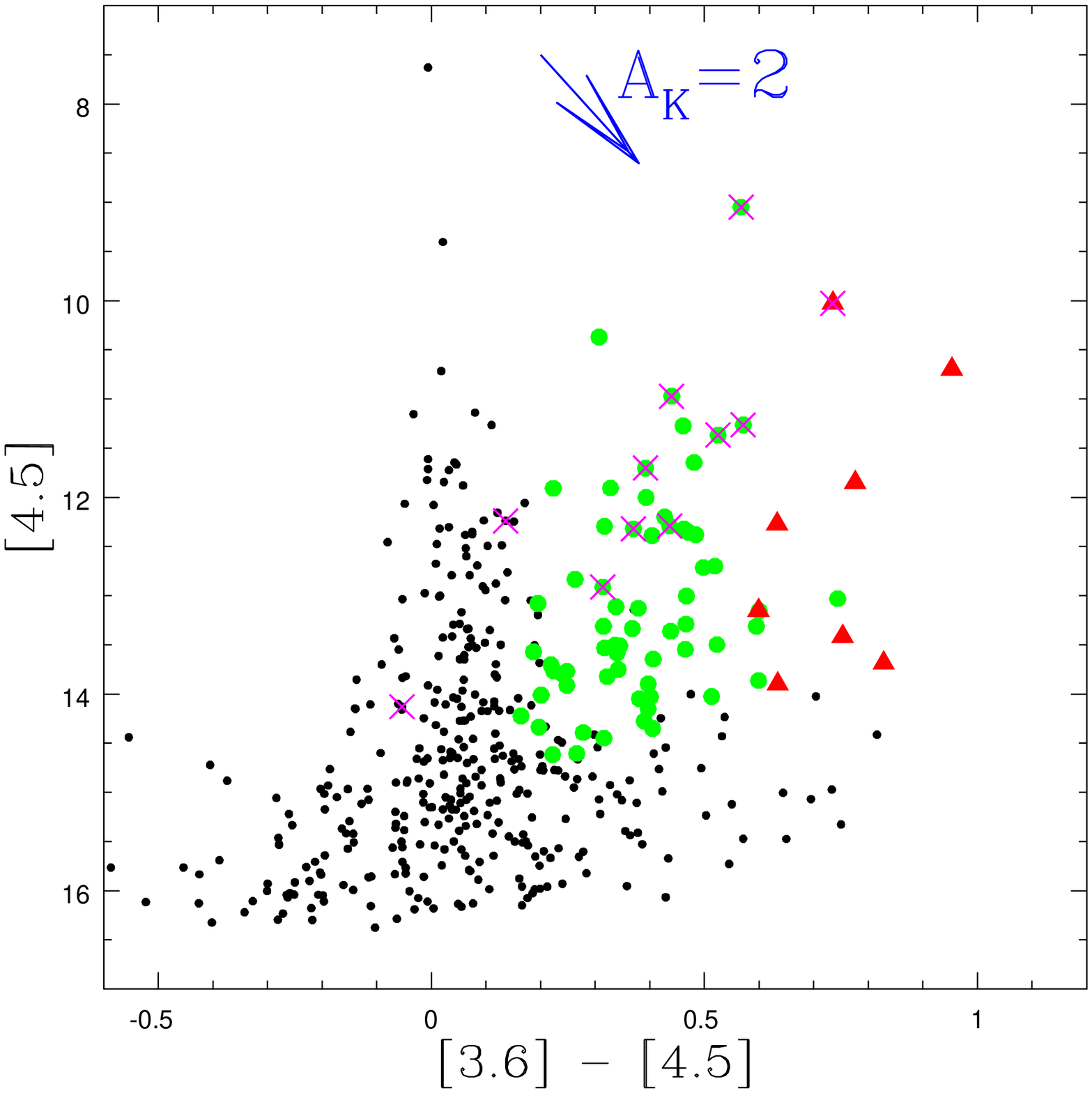}
  \caption{ Colour-magnitude diagrams using IRAC bands along with the candidate YSOs identified. The symbols are same as Fig.~\ref{fig:irac}.}
  \label{fig:mid_ir_cmd}
\end{figure*}

\subsubsection{Selection of YSOs from $H$, $K$, 3.6, 4.5 $\mu$m data}

In order to classify the YSOs,  we used IRAC 3.6 and 4.5 $\mu$m data from {\it Spitzer} and $H$ and $K$ near-IR data. The detection of YSOs based on the {\it Spitzer} data is limited here because  NGC~2282 was observed during the {\it Spitzer} warm mission, which does not provide observations beyond 4.5 $\mu$m. Hence we  missed the longer wavelength IRAC data at 5.6, 8.0 $\mu$m  and MIPS at 24 $\mu$m bands for improved YSO classification. Here we classify  only Class I and II objects and we do not account for Class III objects as they are indistinguishable from field stars based on the data sets used in this study.

We adopted the source classification scheme introduced by Gutermuth et al. (2008, 2009) based on the IR colours. We first dereddened the data  using  our $K$-band extinction map (see sect. 3.3). We minimized the inclusion of extragalactic contamination by applying a simple brightness cut in the dereddened 3.6 $\mu$m photometry, i.e, all the Class I YSOs essentially have $[3.6]_{0}$ $\leq$ 15 mag and all the Class II YSOs essentially have $[3.6]_{0}$ $\leq$ 14.5 mag (Gutermuth et al. 2009).  After removing the contaminants, we identified YSOs from ($[[K]-[3.6]]_{0}, [[3.6]-[4.5]]_{0}$) CC diagram shown in Fig.~\ref{fig:irac}. Thus we identified 84 candidate YSOs , which includes 9 Class I and 75 Class II sources with IR excess emission. In Fig.~\ref{fig:irac}, the Class I and Class II sources are shown as the red triangles and green circles, respectively,  and  a reddening vector for $A_K$ = 2 mag is also plotted by using the reddening law from Flaherty et al. (2007). The $H_\alpha$ emission line sources from slitless spectroscopy are marked with magenta crosses (discussed later).  Fig.~\ref{fig:mid_ir_cmd} shows the CMDs in mid-IR bands for all the uncontaminated sources along with the candidate  YSOs. Our present YSO selection is incomplete as many sources in high nebulous region might have not detected at 3.6 and 4.5 $\mu$m bands as well as due to the detection limits of IRAC observations.

\subsubsection{YSOs from Near-IR Colour-Colour Diagram}

\begin{figure*}
\includegraphics[width=8.0 cm,height=8.0 cm]{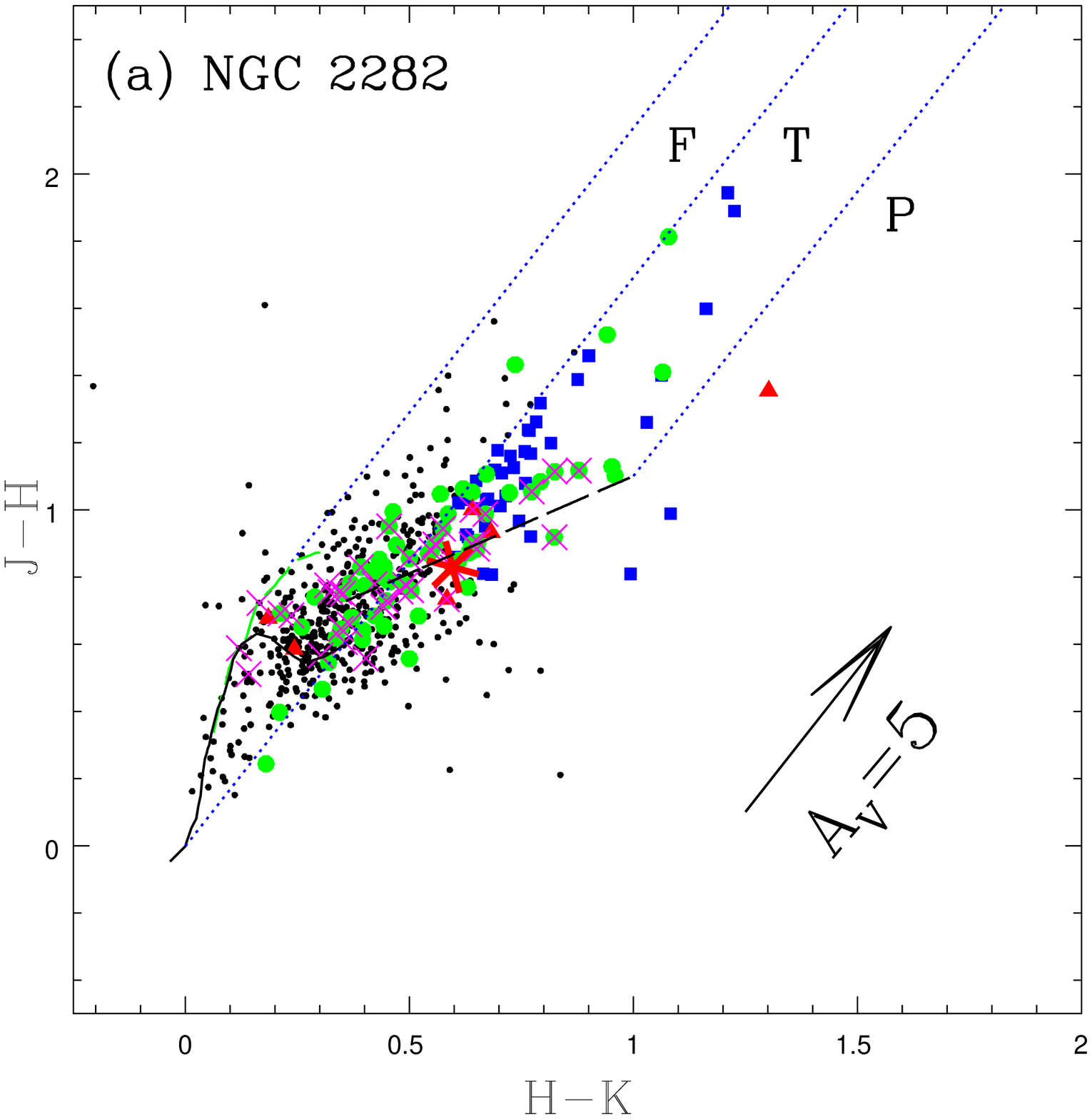}
\includegraphics[width=8.0 cm,height=8.0 cm]{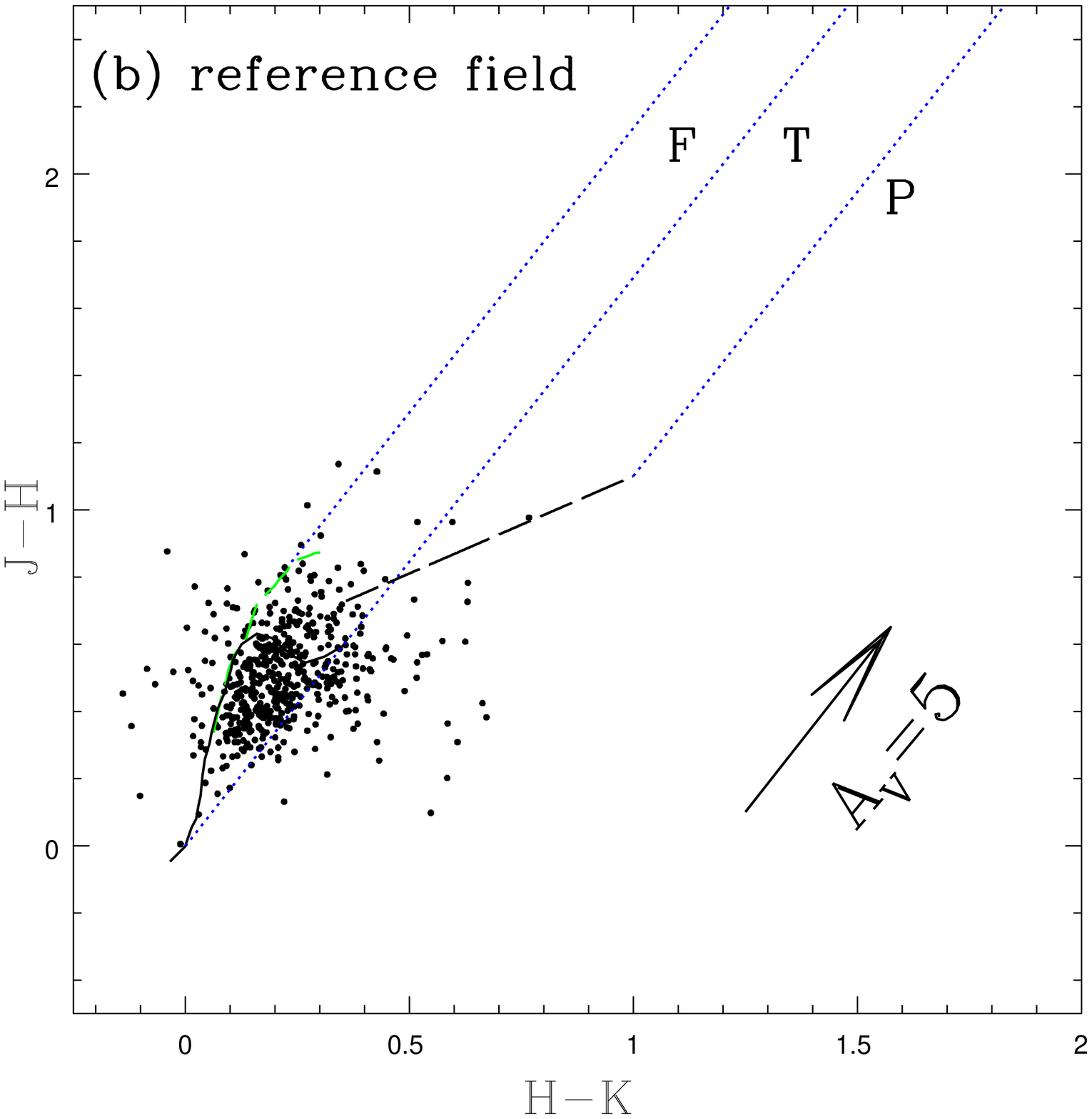}
\caption{CC diagram of (a) NGC 2282 within 3.15$\arcmin$ radius and (b) the reference field of same area. The locus for dwarfs (solid black) and giants (green dashed line) are taken from Bessel $\&$ Brett (1988). The long dashed black line represents the CTTs locus (Meyer et al. 1997) and the small dashed blue lines represent the reddening vectors (Cohen et al. 1981). The reddening vector of visual extinction $A_V$ = 5 mag is also shown. All the Class I and Class II sources are represented with red triangles and green circles, respectively. The blue solid squares are the candidate YSOs identified  from JHK colours. The asterisk mark is the location of Herbig Ae/Be star. The magenta crosses  are the $H_\alpha$ emitting objects detected from slitless spectroscopy and IPHAS photometry (see text).}
  \label{fig:jhk_yso}
\end{figure*}

 Near-IR colour-colour diagram is shown in Fig.~\ref{fig:jhk_yso}a. The black solid and long dashed green line represent the locus of the intrinsic colours of dwarf and giant stars, taken from Bessel \& Brett (1988). The dashed black line represents the locus of the classical T Tauri stars (CTTSs) (Meyer et al. 1997). All the  intrinsic locus and photometric data points are  transformed in to the CIT (California Institute of Technology) system (Elias et al. 1982) using the relations given by Carpenter et al. (2001). The parallel dashed lines represent the interstellar reddening vectors. The slope of the reddening vectors (i.e., $A_J$/$A_V$ = 0.265, $A_H$/$A_V$ = 0.155 and  $A_K$/$A_V$ = 0.090) are taken from Cohen et al. (1981).

We divide the $JHK$ near-IR space into three regions- F, T, and P. The near-IR emission  of stars in `F' region  originate from their discless photosphere. These stars are located between the upper and middle reddening vectors in the  near-IR CC diagram and they are considered to be either field stars or weak-line T Tauri stars (WTTSs)/Class III sources  with no or small near-IR excess. However, it is very difficult to distinguish between WTTSs with small near-IR excess and field stars from only near-IR CC diagram (Ojha et al. 2004).  The near-IR emission of `T' region stars  arise from both photosphere and circumstellar disc ( Lada \& Adams 1992). Majority of these stars are considered to be classical T Tauri stars (CTTSs). All such sources possess accreting optically thick disc (Meyer et al. 1997). The `P' region stars have more near-IR colour excess at K-band, and these stars are thought to have accreting  disc. Some of them might have  envelope around them and they are protostellar in nature  (Rice et al. 2012).

 We also plotted the CC diagram for the reference field sources in Fig.~\ref{fig:jhk_yso}b. The reference field region was chosen for the same area as that of the cluster (radius = 3.15$\arcmin$) and at similar photometric depth. The reference field was selected  $\sim$ 10$\arcmin$  towards the  North of   NGC 2282  ($\alpha_{2000}$ = $06^h46^m51.06^s$ $\delta_{2000}$ = $+01^029^m29.9^s$),  to avoid any superposition with the cluster region.  A comparison of the  reference field with  our cluster region shows  that  almost all the stars in the reference field are confined below $(J-H)$ $\sim $ 0.8 mag and to the left of the middle reddening vector from $(H-K)$ $\sim$ 0.6 mag.  Thus we  assume  that the `T' and `P' region stars are not contaminated by the field stars. Hence those  sources fall in the `T' and `P' regions of the target field could be considered as candidate YSOs. From near-IR colour, we have selected 45 additional YSOs, which are not included in the  previous YSO list (sect. 3.5.1). Only from $JHK$ data it is not possible to distinguish Class I and Class II objects, hence we consider them as candidate YSO sources.

\subsubsection{Selection of H$_\alpha$ Emission Stars from Slitless Spectroscopy and IPHAS Photometry}

Using slitless spectroscopy, $H_\alpha$ emission-line stars were visually identified from their enhancement over the continuum. We identified 16 $H_\alpha$ emitting sources and they are listed in Table~\ref{tab:ysos}.   Of the 16, 14 sources  have $JHK$ data sets. Ten $H_\alpha$ emission stars are detected as YSOs from $H$, $K$, 3.6 and 4.5 $\mu$m colours (1 Class I, 9 Class II). All the $H_\alpha$ emission stars are plotted in the near-IR colour-colour diagram shown in \ref{fig:jhk_yso}a. The star ID 8 (see Table~\ref{tab:spec}), which is classified as a B0.5 Ve star in the optical slit-spectroscopy with strong $H_\alpha$ emission, is also detected with the slitless spectroscopy and falling in the `P' region of $JHK$ CC diagram (See Fig.~\ref{fig:jhk_yso}a).

\begin{figure}
\includegraphics[width=8.0 cm,height=8.0 cm]{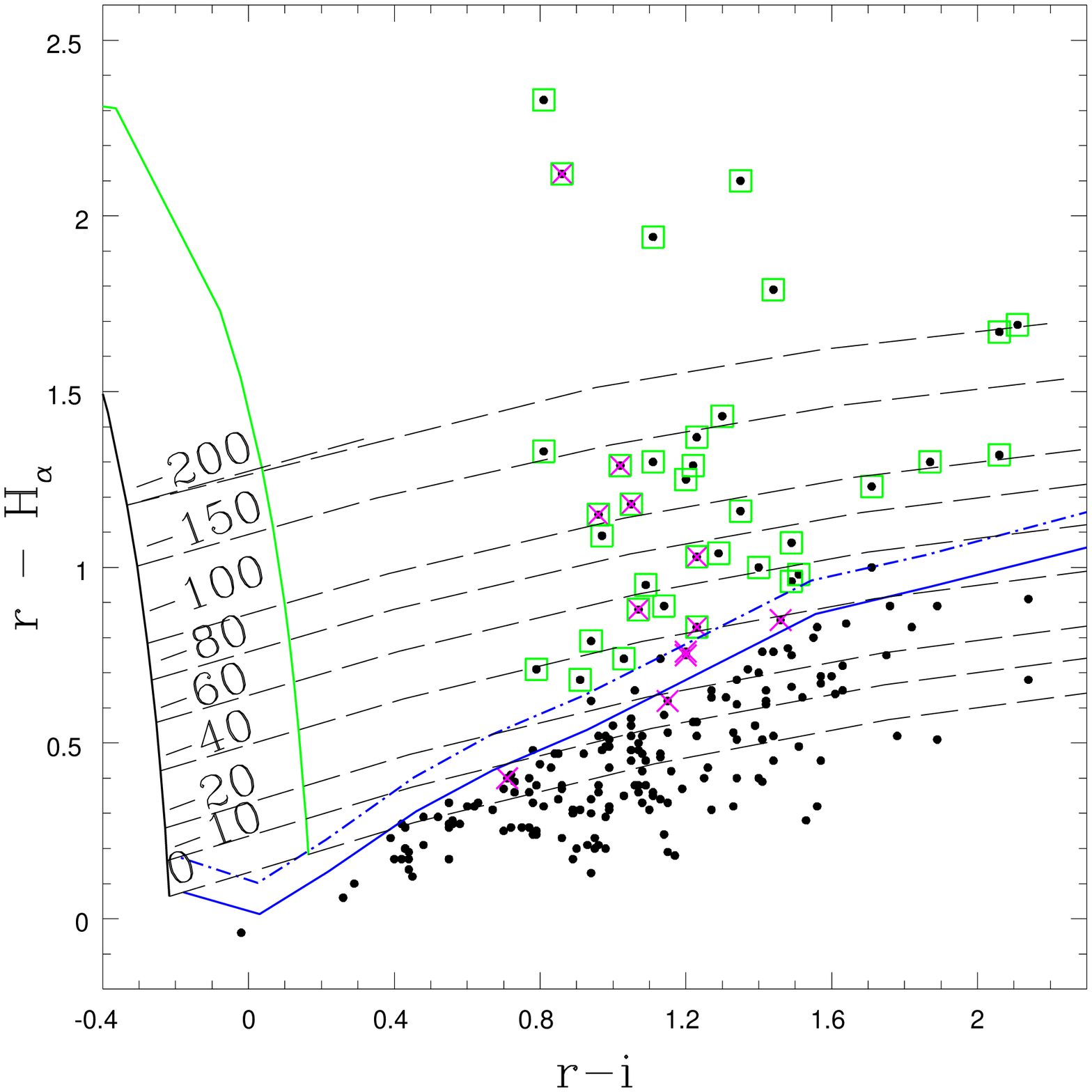}
  \caption{$r-i/r-H_\alpha$ CC diagram for all the sources detected in IPHAS photometry towards NGC 2282. The magenta crosses and green boxes are  the  H$_\alpha$ emission sources detected   from   slitless spectroscopy and IPHAS photometry, respectively. See text for increasing levels of $H_\alpha$ emission tracks, unreddened continuum and locus of main-sequence.} 
 
  \label{fig:barentsen}
\end{figure}

Fig.~\ref{fig:barentsen} presents the IPHAS ($r-i/r-H_\alpha$) CC diagram towards NGC 2282.  Black filled circles are the sources brighter than $r$ $<$ 20 with photometric uncertainty of $<$ 0.1 mag in IPHAS DR2. The identified sources are marked in Fig.~\ref{fig:barentsen}. Two nearly vertical black and green lines represent the trend for an unreddened Rayleigh-Jeans continuum  and the case of an unreddened optically thick disc accretion continuum, respectively (Barentsen et al. 2014). The black broken lines are the predicted lines of constant net emission EW. The solid and broken blue lines indicate the locus of unreddened main-sequence and that of the main-sequence stars having an H$_\alpha$ emission-line strength of $-$10 \AA $~$ EW, respectively. The main-sequence emission line with EW $-$10 \AA $~$ is chosen as CTTS threshold for H$_\alpha$ emission stars. However, it is difficult to confirm CTTSs solely on the basis of IPHAS photometry (Barentsen et al. 2014). Other possible H$_\alpha$ emission objects are evolved massive stars (e.g. Be stars, Wolf-Rayet stars, luminous blue variables), evolved intermediate-mass stars (e.g. Mira Variables, unresolved planetary nebulae) and interacting binaries (e.g. cataclysmic variables, symbiotic stars). (Barentsen et al. 2011; Corradi et al. 2008).

We selected 44 sources as $H_\alpha$ emitting stars, which are located above the 3$\sigma$ confidence level  from CTTSs threshold as mentioned above. Out of 16 $H_\alpha$ emitting sources detected from slitless spectroscopy, 15 have IPHAS photometry, but only 10 satisfy the conditions for $H_\alpha$ emitting stars using IPHAS photometry. This difference could be due to the variable $H_\alpha$ emission activities in PMS sources as well as due to the different detection limits for these two observations. Thus we have selected 50 $H_\alpha$ emitting sources from slitless spectroscopy and IPHAS photometry. Among those 50 $H_\alpha$ emitting sources, 24 are classified  as Class II and 3 are classified as Class I sources. Including these $H_\alpha$ emission sources, and the YSOs selected from mid-IR and near-IR colours in sect. 3.5.1 and sect. 3.5.2, we have 152 candidate YSOs in NGC 2282 region, and the details  are presented in Table~\ref{tab:ysos}.

\begin{figure*}
\includegraphics[width=8.0 cm,height=8.0 cm]{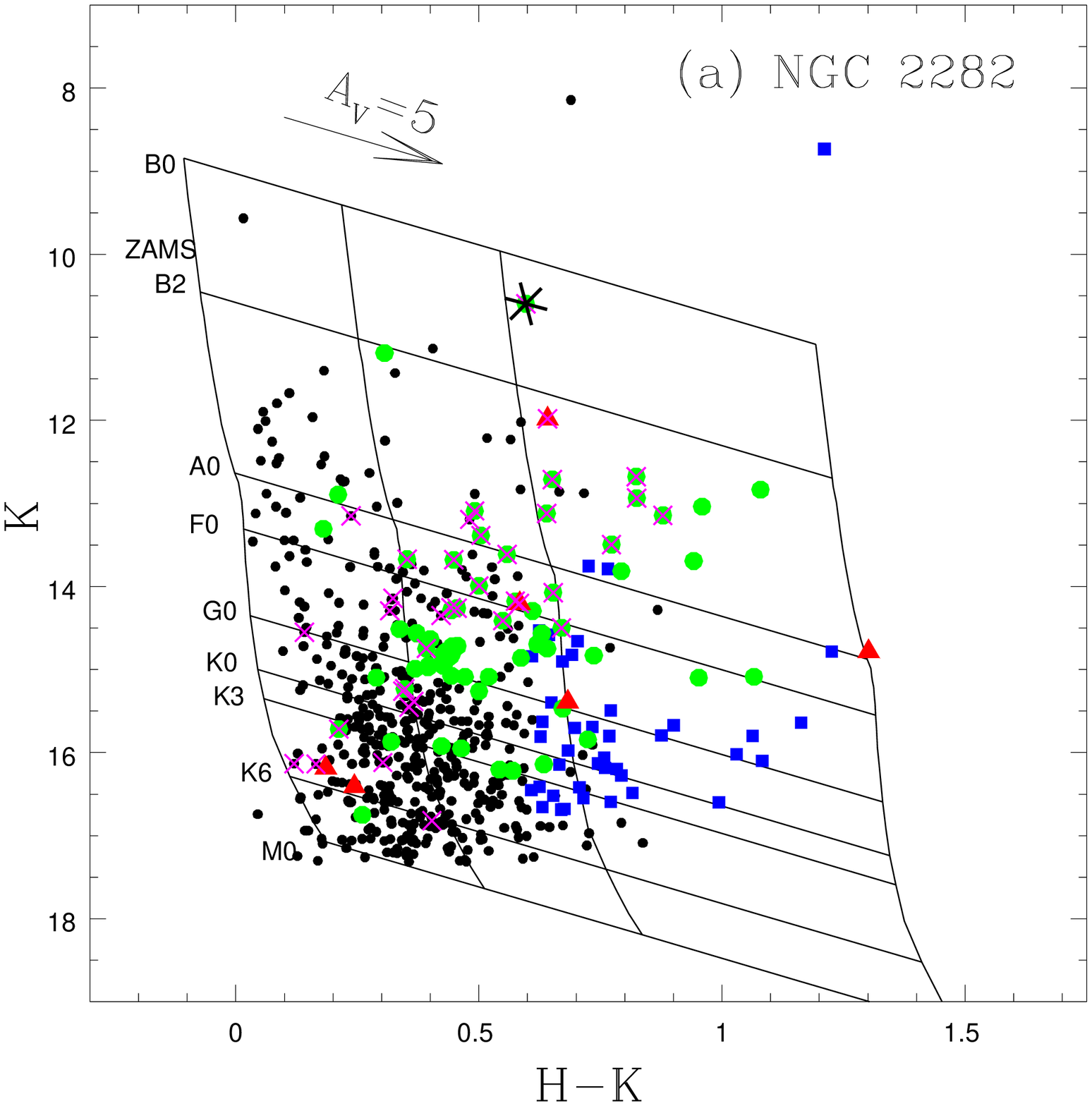}
\includegraphics[width=8.0 cm,height=8.0 cm]{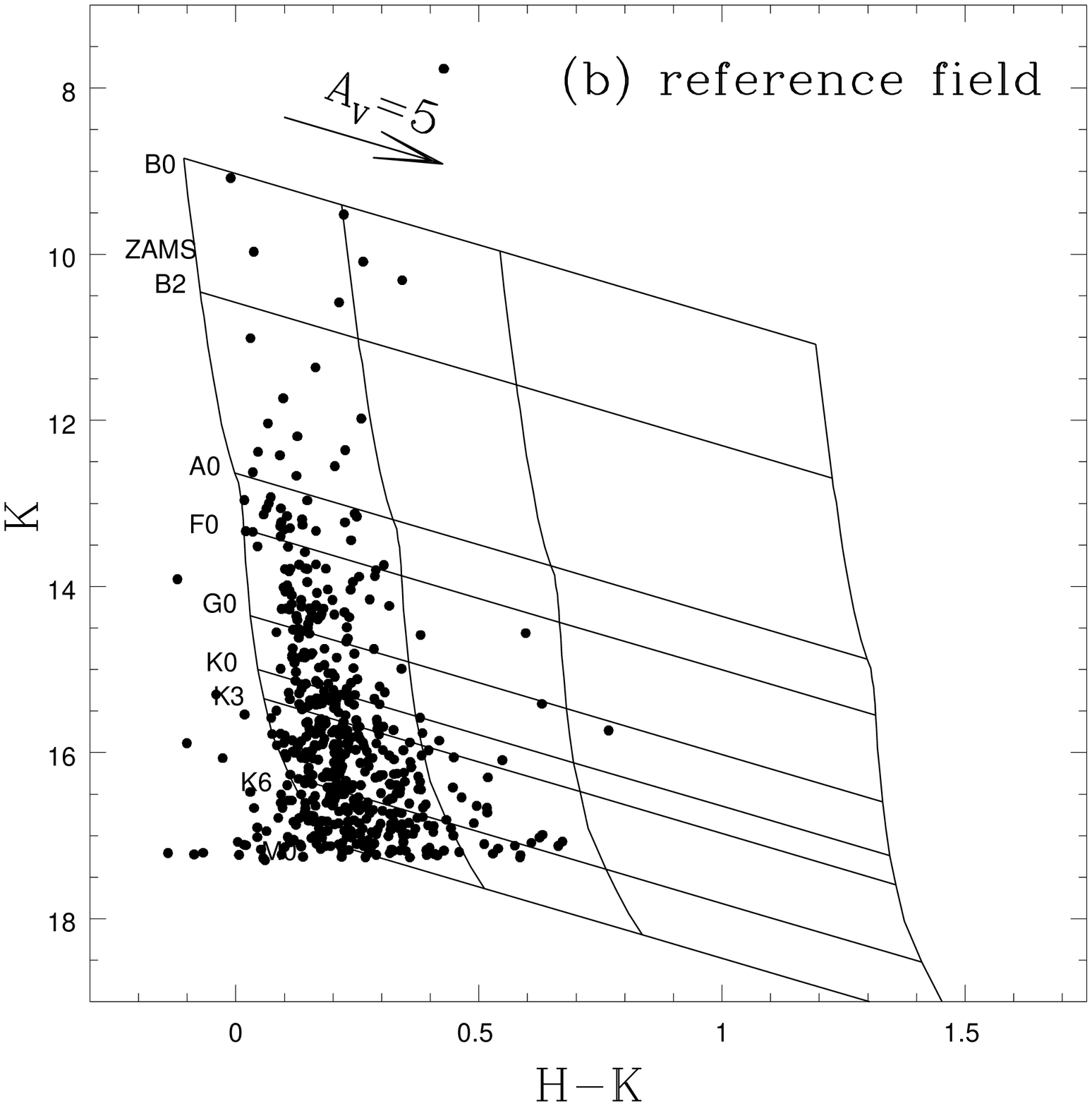}
  \caption{(a) Near-IR CM diagram of the stars within 3.15$\arcmin$ radius of NGC 2282. The nearly vertical solid lines are the loci of ZAMS stars at 1.65 kpc reddened by $A_V$ = 0, 5, 10, 20 mag.  The slanting horizontal lines are the reddening vectors for corresponding spectral classes. All the Class I and Class II sources  determined from the $H$, $K$, 3.6 and 4.5 $\mu$m data are represented with red triangles and green circles, respectively. The blue solid squares are the candidate  YSOs selected from $JHK$ colours. The magenta crosses are the $H_\alpha$ emitting objects detected from the slitless spectroscopy and IPHAS photometry.  The asterisk is the location of Herbig Ae/Be star. The  (b) CM diagram for the reference field.}
  \label{fig:khk}
\end{figure*}

\renewcommand{\tabcolsep}{2.0pt} 
\begin{table*}
\centering
\caption{Photometric catalog of candidate YSOs in NGC 2282 (Class I, Class II, candidate YSOs detected from $JHK$ colours and $H_\alpha$ emission sources). The complete table is available in  electronic version.}
\tiny
\label{tab:ysos}
\begin{tabular}{|r|r|r|r|r|r|r|r|r|r|r|r|||||||||}
\hline \multicolumn{1}{c}{ID} & \multicolumn{1}{c}{$\alpha_{2000}$} & \multicolumn{1}{c}{$\delta_{2000}$} & \multicolumn{1}{c}{$V$} & \multicolumn{1}{c}{$B-V$} & \multicolumn{1}{c}{$V-I$} & \multicolumn{1}{c}{$J$} & \multicolumn{1}{c}{$H$} & \multicolumn{1}{c}{$K$} & \multicolumn{1}{c}{3.6 $\mu$m} & \multicolumn{1}{c}{4.5 $\mu$m} & \multicolumn{1}{c}{H$_\alpha$-emission}\\ 

\multicolumn{1}{c}{} & \multicolumn{1}{c}{(deg)} & \multicolumn{1}{c}{(deg)} & \multicolumn{1}{c}{(mag)} & \multicolumn{1}{c}{(mag)} & \multicolumn{1}{c}{(mag)} & \multicolumn{1}{c}{(mag)} & \multicolumn{1}{c}{(mag)} & \multicolumn{1}{c}{(mag)} & \multicolumn{1}{c}{(mag)} & \multicolumn{1}{c}{(mag)}& \multicolumn{1}{c}{(YES/NO)} \\ \hline
\multicolumn{12}{c}{Class I sources}
\\
\hline 
  54  &  101.723024  &  1.310400  &  17.447 $\pm$  0.005  &   1.323  $\pm$  0.018  &    2.195  $\pm$  0.005 &   13.677  $\pm$  0.026  &  12.644  $\pm$  0.026  &  11.958  $\pm$   0.027  &  10.761  $\pm$  0.003  &  10.026  $\pm$  0.002 & YES\\
  310  &  101.715435  &  1.271265   &  21.316 $\pm$  0.039  &   1.434  $\pm$  0.248  &    2.224  $\pm$  0.046 &   17.144  $\pm$  0.024  &  16.131  $\pm$  0.016  &  15.420  $\pm$   0.017  &  13.749  $\pm$  0.013  &  13.150  $\pm$  0.011 & NO\\
  313  &  101.728048  &  1.284664  &  20.794 $\pm$  0.033  &   1.692  $\pm$  0.188  &    2.195  $\pm$  0.036 &   17.129  $\pm$  0.024  &  16.393  $\pm$  0.021  &  16.200  $\pm$   0.034  &  14.514  $\pm$  0.021  &  13.686  $\pm$  0.015 &NO \\
  644  &  101.768647  &  1.344268  &  18.882 $\pm$  0.008  &   1.331  $\pm$  0.026  &    1.844  $\pm$  0.010 &   16.088  $\pm$  0.010  &  14.724  $\pm$  0.005  &  13.497  $\pm$   0.003  &  12.051  $\pm$  0.004  &  11.494  $\pm$  0.004 & YES\\
 1873  &  101.729427  &  1.289632  &         $...$   $~~~~~~$      &         $...$   $~~~~~~$        &          $...$    $~~~~~~$      &   17.665  $\pm$  0.038  &  16.193  $\pm$  0.017  &  14.837  $\pm$   0.010  &  12.903  $\pm$  0.008  &  12.270  $\pm$  0.006 & NO\\
 1972  &  101.751821  &  1.318276  &         $...$  $~~~~~~$        &         $...$  $~~~~~~$          &          $...$   $~~~~~~$        &          $...$   $~~~~~~$         &  17.810  $\pm$  0.076  &  16.137  $\pm$   0.033  &  14.167  $\pm$  0.013  &  13.414  $\pm$  0.012 & NO\\
 2304  &  101.717280  &  1.315665  &         $...$ $~~~~~~$         &         $...$  $~~~~~~$          &          $...$         $~~~~~~$  &   17.309  $\pm$  0.028  &  16.674  $\pm$  0.027  &  16.420  $\pm$   0.043  &  14.532  $\pm$  0.046  &  13.898  $\pm$  0.019 & NO\\
 2329  &  101.712510  &  1.313155  &         $...$   $~~~~~~$       &         $...$   $~~~~~~$         &          $...$   $~~~~~~$        &   15.636  $\pm$  0.007  &  14.838  $\pm$  0.005  &  14.230  $\pm$   0.006  &  12.626  $\pm$  0.007  &  11.850  $\pm$  0.005 &  YES\\
 2493  &  101.734794  &  1.342232   &         $...$    $~~~~~~$      &         $...$     $~~~~~~$       &          $...$   $~~~~~~$        &   19.570  $\pm$  0.220  &  16.985  $\pm$  0.036  &  14.804  $\pm$   0.010  &  11.650  $\pm$  0.004  &  10.697  $\pm$  0.003 & NO\\
\hline 
\multicolumn{12}{c}{Class II sources}
\\
\hline 
    1  &  101.709618  &  1.276534  &  14.674 $\pm$  0.008  &   0.933  $\pm$  0.007  &    1.438  $\pm$  0.004 &   11.964  $\pm$  0.020  &  11.506  $\pm$  0.029  &  11.164  $\pm$   0.034  &  10.678  $\pm$  0.002  &  10.371  $\pm$  0.003 & NO \\
    8  &  101.735121  &  1.277934  &  15.934 $\pm$  0.013  &   1.037  $\pm$  0.010  &    1.841  $\pm$  0.007 &   12.061  $\pm$  0.023  &  11.212  $\pm$  0.020  &  10.572  $\pm$   0.023  &   9.616  $\pm$  0.003  &   9.049  $\pm$  0.002 & YES \\
   71  &  101.705376  &  1.348143  &  15.176 $\pm$  0.003  &   0.660  $\pm$  0.003  &    0.887  $\pm$  0.005 &   13.715  $\pm$  0.026  &  13.496  $\pm$  0.035  &  13.283  $\pm$   0.033  &  12.863  $\pm$  0.006  &  12.379  $\pm$  0.007 & NO\\
   72  &  101.726013  &  1.361580  &  15.389 $\pm$  0.003  &   0.752  $\pm$  0.004  &    1.070  $\pm$  0.003 &   13.498  $\pm$  0.021  &  13.114  $\pm$  0.032  &  12.870  $\pm$   0.034  &  12.233  $\pm$  0.005  &  11.905  $\pm$  0.005 & NO\\
   97  &  101.804643  &  1.326651  &  13.103 $\pm$  0.005  &   0.461  $\pm$  0.005  &    0.878  $\pm$  0.004 &   11.618  $\pm$  0.019  &  11.209  $\pm$  0.020  &  10.952  $\pm$   0.021  &  10.420  $\pm$  0.002  &  10.225  $\pm$  0.003 & NO\\
  101  &  101.775393  &  1.354269  &  15.659 $\pm$  0.003  &   1.161  $\pm$  0.004  &    1.501  $\pm$  0.004 &   13.017  $\pm$  0.023  &  12.240  $\pm$  0.024  &  11.840  $\pm$   0.023  &  11.168  $\pm$  0.003  &  10.974  $\pm$  0.003 & NO\\
  289  &  101.748231  &  1.277375  &  20.235 $\pm$  0.018  &   1.594  $\pm$  0.123  &    2.514  $\pm$  0.014 &   15.275  $\pm$  0.005  &  14.077  $\pm$  0.003  &  13.078  $\pm$   0.002  &  12.126  $\pm$  0.005  &  11.645  $\pm$  0.005 & NO\\
  298  &  101.715135  &  1.273663  &  20.994 $\pm$  0.037  &   1.632  $\pm$  0.217  &    2.548  $\pm$  0.031 &   16.182  $\pm$  0.010  &  15.294  $\pm$  0.008  &  14.851  $\pm$   0.010  &  14.212  $\pm$  0.018  &  14.011  $\pm$  0.018 & NO\\
  307  &  101.716418  &  1.277292  &  21.022 $\pm$  0.034  &   1.443  $\pm$  0.182  &    2.505  $\pm$  0.030 &   16.320  $\pm$  0.012  &  15.423  $\pm$  0.009  &  14.978  $\pm$   0.012  &  14.387  $\pm$  0.020  &  14.223  $\pm$  0.021 & NO\\
  321  &  101.719373  &  1.292577  &  19.190 $\pm$  0.013  &   1.621  $\pm$  0.050  &    2.329  $\pm$  0.009 &   15.042  $\pm$  0.004  &  13.832  $\pm$  0.002  &  12.973  $\pm$   0.002  &  11.412  $\pm$  0.003  &  10.972  $\pm$  0.003 & NO\\
  363  &  101.759568  &  1.306980  &  21.575 $\pm$  0.061  &         $...$   $~~~~~~$         &    2.862  $\pm$  0.046 &   15.855  $\pm$  0.008  &  14.678  $\pm$  0.005  &  13.852  $\pm$   0.005  &  12.626  $\pm$  0.006  &  12.199  $\pm$  0.006 & NO \\
  364  &  101.762999  &  1.308987  &  20.603 $\pm$  0.056  &         $...$    $~~~~~~$        &    2.610  $\pm$  0.029 &   15.653  $\pm$  0.007  &  14.770  $\pm$  0.005  &  14.308  $\pm$   0.007  &  13.212  $\pm$  0.009  &  12.714  $\pm$  0.008 & NO\\

\hline
\hline\end{tabular}
\end{table*}

\section{Discussion}
\subsection{Near-IR Colour-Magnitude Diagram }

The near-IR CMD is a useful tool for understanding the nature of YSOs in the embedded star-forming regions. The near-IR CMD ($K$ vs $(H-K)$) for all the stars detected towards NGC 2282 cluster is shown in Fig.~\ref{fig:khk}a. The identified all sources are marked. A reference field of same area and at similar photometric depth is also shown in Fig.~\ref{fig:khk}b. The nearly vertical solid lines are the loci of ZAMS reddened by visual extinction of $A_V$ = 0, 5, 10, 20 magnitude, respectively and corrected for cluster distance of 1.65 kpc. The slanting horizontal lines represent the reddening vectors of the corresponding  spectral type made from Flaherty et al. (2007). The membership of any YSO relies on the fact that they are mainly found in the cluster area rather than in the surrounding field. We considered only those YSOs within the cluster area (radius = 3.15$\arcmin$) to determine the cluster parameters. In the CMD given in Fig.~\ref{fig:khk}a, we can see that majority of the YSOs are located within B2 to K6 spectral type. Majority of $H_\alpha$ emission objects fall within B0 to F0 spectral type. A Herbig Ae/Be type star, determined as B0.5 from spectroscopic observations, falls close to B0 vector with $A_V$ $\sim$ 10 mag. This  estimate matches well with the spectroscopic observations (sect. 3.2). The YSOs, determined from IR CC diagrams, share IR space with many unclassified sources marked as black dots. These might be the background sources or the weak line T Tauri sources, which are not included in our YSO survey. Spectroscopic observations are necessary for the confirmation of their membership.

\subsection{Optical Colour-Magnitude Diagram of YSOs}

\begin{figure}
\includegraphics[width=8.0 cm,height=8.0 cm]{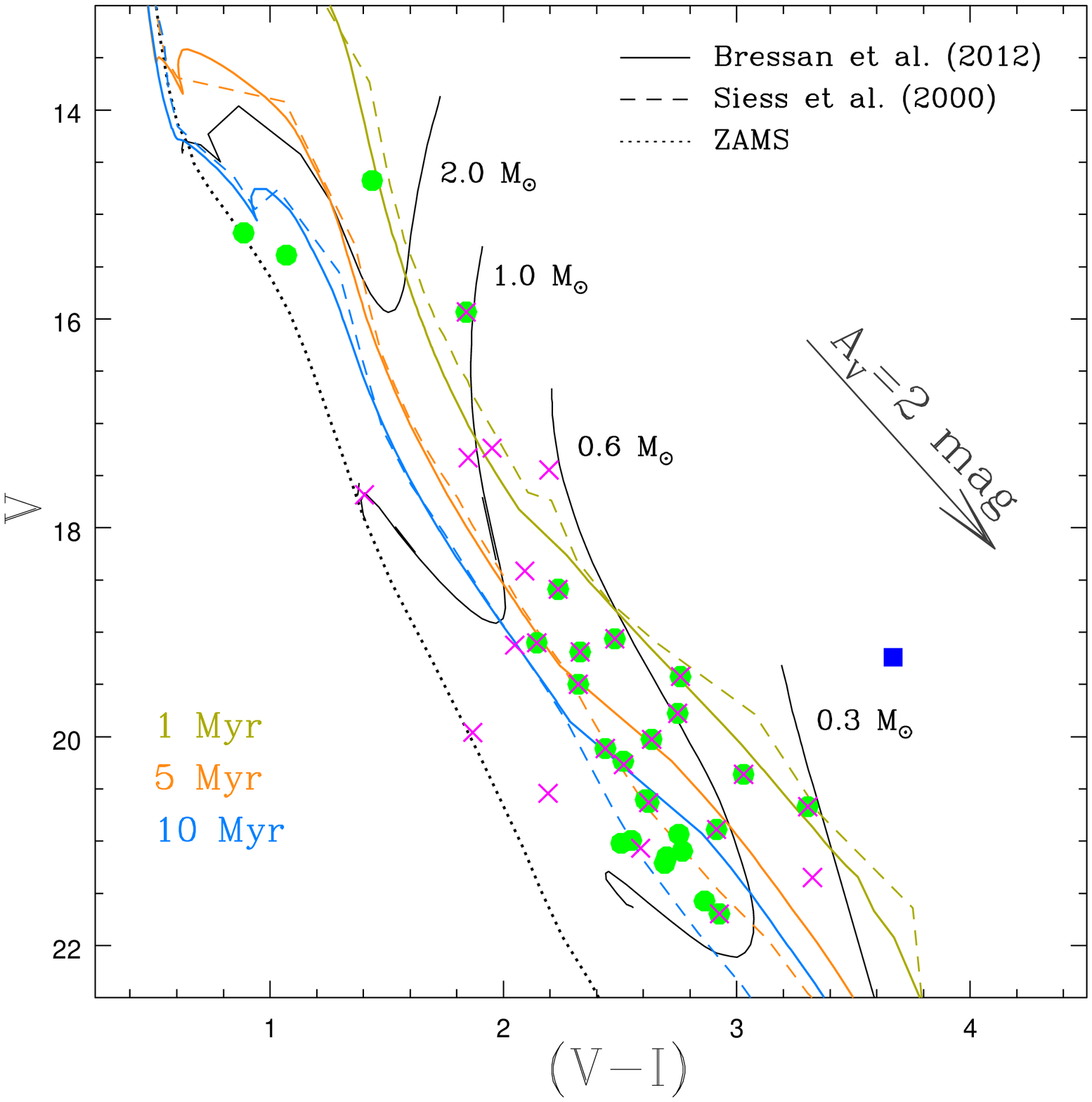}
  \caption{$V/(V-I)$ CMD for the Class II sources, candidate YSOs from $JHK$ colours and $H_\alpha$ emitting sources within NGC 2282 cluster. All the symbols are same as in Fig.~\ref{fig:jhk_yso}. The dotted curve is the locus of ZAMS from Girardi et al. (2002), solid curves are the PMS isochrones of age 1.0, 5.0 and 10.0 Myr, respectively, and the thin black solid curves are the evolutionary tracks for various mass bins from Bressan et al. (2012). The long dashed curves are PMS isochrones of age 1.0, 5.0 and 10.0 Myr taken from Siess et al. (2000). All the isochrones and tracks are corrected for the distance and reddening. }
  \label{fig:viv_yso}
\end{figure}

 An Optical CMD, $V$ vs $(V-I)$ of YSOs is plotted in Fig.~\ref{fig:viv_yso}. It is an important tool to estimate the approximate ages and masses of YSOs. The solid curve in Fig.~\ref{fig:viv_yso} represents the  ZAMS, taken from Girardi et al. (2002) corrected for the cluster distance 1.65~kpc and reddening of $E(B-V)$ = 0.52 mag ($E(V-I)$ = 0.65 mag) (see sect. 3.3 \& 3.4). We have used the PMS isochrone and evolutionary tracks of Bressan et al.(2012) to determine the ages and masses of the YSOs. The PMS isochrones for Siess et al. (2000) is also plotted for comparison. The ages and masses of  YSOs have been estimated by comparing their locations on the CMD with PMS isochrones of various ages after correcting for the distance and extinction. Since the reddening vector is nearly parallel to the isochrones, a small extinction variation would not have much effect on the age estimation  of YSOs.    

 We have compiled  $VI$ photometry for the YSOs, and compare their positions on the CMD to theoretical model isochrones. The CMD positions of YSOs seem to be adequately fit between 1$-$10 Myr. Different models at low-mass end differs significantly as we can see in Fig.~\ref{fig:viv_yso}. The average age of the YSOs seem to be $\sim$ 2-5 Myr, which we considered as the average age of the cluster.

\subsection{Mass Distribution}

\begin{figure}
\includegraphics[width=8.0 cm,height=8.0 cm]{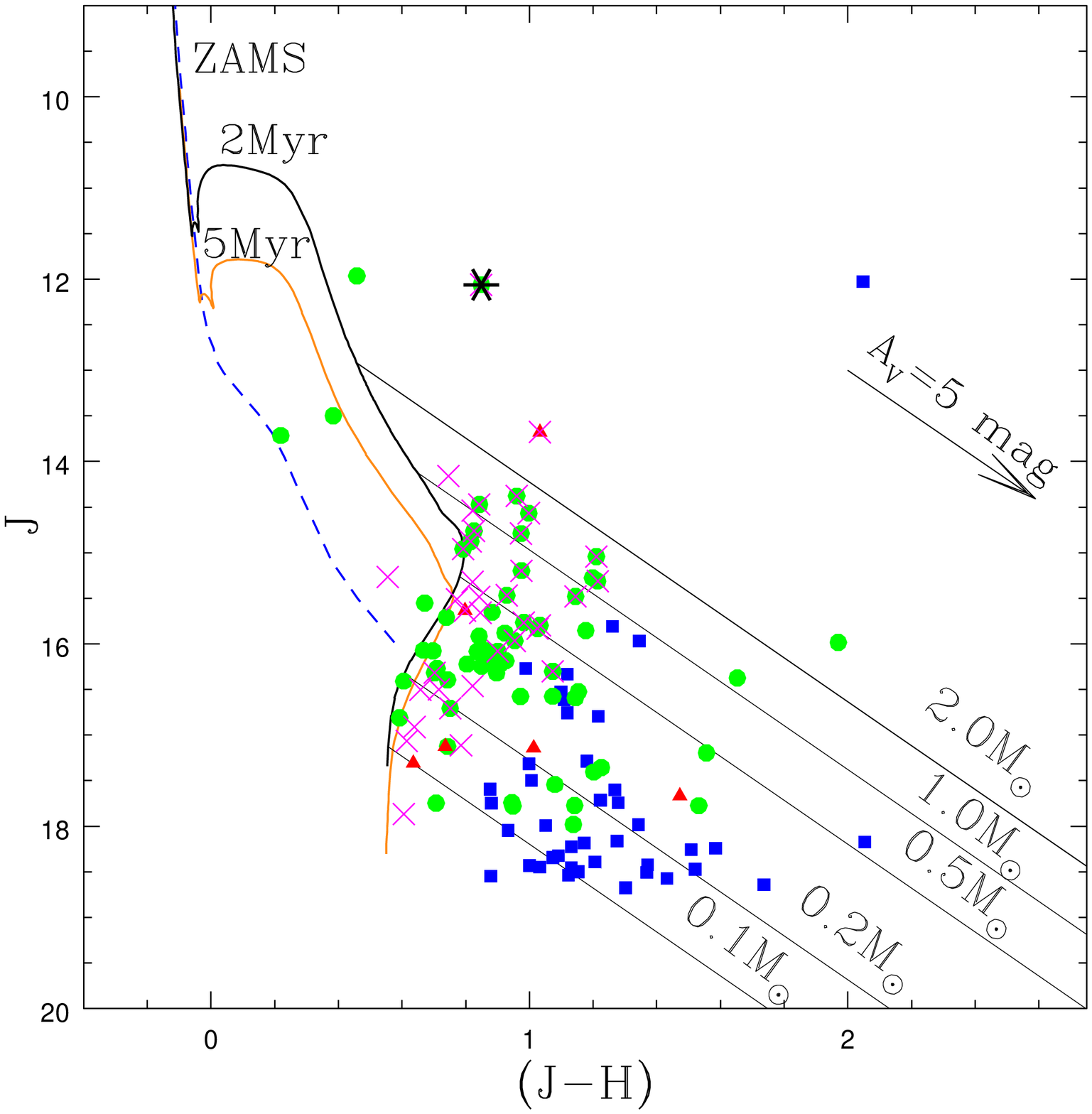}
  \caption{$J$ vs $(J-H)$ diagram for Class I, Class II sources, candidate  YSOs detected from $JHK$ colours and the $H_\alpha$ emitting sources. All the symbols are same as in Fig.~\ref{fig:jhk_yso}. Herbig AeBe star (asterisk mark) is also shown. The blue dashed curve is the locus of the ZAMS from Girardi et al. (2002). The solid curves are the PMS isochrones of age  2.0 and 5.0 Myr, respectively, from Bressan et al. (2012). The slanting solid lines are the reddening vectors corresponding to 0.1, 0.2, 0.5, 1.0 and 2.0 M$_\odot$.}
  \label{fig:jhj_yso}
\end{figure}

Since YSOs show excess emission at longer wavelengths, $K/H-K$ CM diagram is not a suitable tool to understand the mass distribution of YSOs. We used $J$ vs $(J-H)$ CM diagram to reduce the effect of excess emission. Fig.~\ref{fig:jhj_yso} shows the  $J/(J-H)$ CM diagram for all the YSOs detected from IR colour-colour diagrams (Fig.~\ref{fig:irac}, \ref{fig:jhk_yso}a, \ref{fig:barentsen}). The ZAMS of Giradi et al. (2002) corrected for the cluster distance of 1.65 kpc is used for comparison in Fig.~\ref{fig:jhj_yso}. The PMS isochrones for 2.0 Myr and 5.0 Myr are taken from Bressan et al. (2012).

It is apparent from Fig.~\ref{fig:jhj_yso} that majority of the candidate YSOs have masses  $\sim$ 0.1$-$2 M$_\odot$. Few candidate YSOs (e.g. ID = 8, asterisk mark) seem to be  apparently more massive and we can not determine their parameters properly as they are highly embedded. Since the low-mass end of the isochrones are very close to each other, a change of $\sim$ 1$-$2 Myr in age would not change the masses drastically. Thus we used a representative age of those YSOs as 2 Myr. However, such mass estimation could be associated with several errors such as presence of binary companions, circumstellar envelope, variable stars and other unknown excess emissions etc. (Samal et al. 2014).

\subsection{Spatial Distribution of YSOs}

\begin{figure*}
\includegraphics[width=16.0 cm,height=14.0cm]{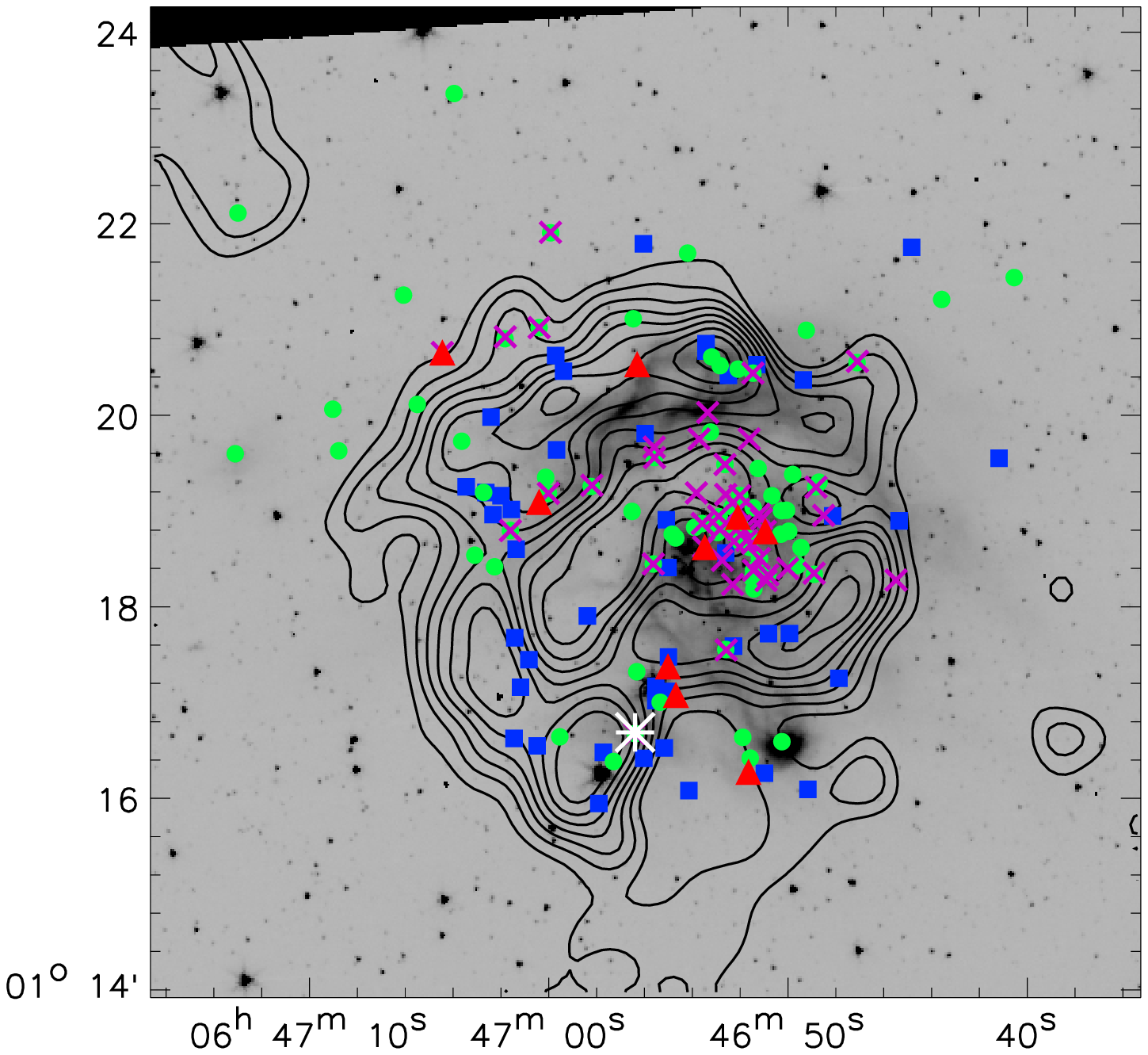}
  \caption{Spatial distribution of Class I (red triangles), Class II (green circles), candidate YSOs from $JHK$ colours (blue squares) overlaid on IRAC 3.6 $\mu$m image. The magenta crosses indicate the H$_\alpha$ emission line sources. The location of Herbig AeBe star (white asterisk) is also shown. The  contours are plotted for different $A_K$ values from 0.32 to 0.87 mag.}
  \label{fig:spatial}
\end{figure*}

Spatial distribution of YSOs in the young clusters traces the star-forming history of that cloud. Fig.~\ref{fig:spatial} shows the spatial distribution of the candidate YSOs within NGC 2282 (red triangles: Class I; green circles: Class II; Blue squares: candidate PMS stars and magenta crosses: $H_\alpha$ emission sources) overlaid on the IRAC 3.6 $\mu$m mosaic image. From Fig.~\ref{fig:spatial}, it is apparent that majority of YSOs are concentrated at the core region of the cluster (see sect. 3.1). Majority ($\sim$ 70\%) of the $H_\alpha$ emitters including IR excess emitters are located in the core region, and some are scattered in the northern side of the cloud.  A secondary peak is seen towards the South-Eastern part of the cluster, which  harbors a Herbig B0.5 Ve star. An interesting arc is visible from East to West through North. No significant population of YSOs are seen in the  Western part of the cloud. While a significant and scattered population towards North-East part of the cloud, particularly Class II, are visible. The extinction map is overplotted in Fig.~\ref{fig:spatial} and it shows that majority of the YSOs are crowded mainly in the central (low extinction) areas of the region.

\subsection{Disc Fraction and Age of the Cluster}

YSOs are surrounded  by circumstellar disc of  gas and dust and the  Near-IR excess emission  originates from the disc (Lada \& Adams 1992). The fraction of sources with excess over the entire number of sources would give an approximate age of the cluster. The disc fraction remains very high ($\geq$ 80 \%) at early stages ($\sim$ 0.3 Myr) of the clusters and decreases with the increasing age (Haisch et al. 2001). The disc lasts for small time scale of about $\sim$ 3$-$15 Myr (Strom et al. 1989;  Lada \& Lada 1995; Haisch et al. 2001; Hillenbrand 2002;  Hern\'{a}ndez et al. 2007) and disc fraction reaches to one-half in $\leq$ 3 Myr time scale (Haisch et al. 2001). The NGC 2282 cluster is physically associated with molecular cloud, which indicates that the cluster has an age of $<$ 10 Myr (Leisawitz et al. 1989). 

We have detected total 1050 objects from our observations within 3.15 arcmin of the cluster radius.  While in similar depth of observations and area on reference field, we have detected 856 stars. After removing the field star contribution, the number of objects associated the cluster area would be $\sim$ 203.  The number of  candidate  YSOs from $H$, $K$, 3.6, 4.5 $\mu$m data  is  72 in the cluster area, while from  $J$, $H$, $K$ data the number  is 40. All disc bearing H$_\alpha$ stars are detected  as the IR excesses sources. Hence, the total number of IR excess sources within NGC 2282 is 112. Thus we estimated the disc fraction of NGC 2282 as  $\sim$ 58\% $\pm$ 6\%. If we consider the IR excess sources from mid-IR data only, the disc fraction is $\sim$  37\% $\pm$ 5\%. This disc fraction estimated in NGC~2282 is significantly larger than that of Horner et al. (1997). This is mainly because  the detection limit of current study is much deeper than the former one. From disc-fraction, the age of the cluster could be in the range $\sim$ 2$-$5 Myr, which is in agreement with the CMD analysis.

\section{Summary and Conclusions}

In this paper, we have presented  multiwavelength studies of a young cluster NGC 2282 in Monoceros constellation, using  deep optical $BVI$ observations complimented with the archival  data sets from IPHAS,  UKIDSS,  2MASS  and mid-IR data from {\it Spitzer} 3.6 and  4.5 $\mu$m. We have also used the spectroscopy observations of 8 bright sources in the cluster region. 
The main results are summarized as follows:

\begin{enumerate}

 \item We have analysed the stellar surface density distribution of  $K$-band data using nearest neighborhood technique. The radius of the cluster has been estimated to be $\sim$ 3.15$\arcmin$ from the semi-major axis of the outer most elliptical contour.

\item  We have estimated  the spectral types  and membership status of 8 bright sources located inside the cluster area using conspicuous lines and comparison of equivalent widths. We have identified three early B-type members in the cluster. Among these B-type massive members, HD 289120, a B2V type star was classified earlier, and two stars ( a Herbig Ae/Be star and a B5 V ) are classified for the first time in this work. We have estimated the distance to the cluster as $\sim$ 1.65 kpc from spectrophotometric analysis of those massive members.

\item The $K$-band extinction map is estimated from $(H-K)$ colours using nearest neighborhood technique, and the mean extinction within the cluster area is found to be $A_V$ $\sim$ 3.9 mag. The extinction within the cluster region seem to be non-uniform.

\item From slitless spectroscopy, we have identified 16 $H_\alpha$ emission line stars. Another 34 $H_\alpha$ emission line stars are identified from IPHAS data, totaling 50 $H_\alpha$ emission line stars towards the region. 

\item  Using Gutermuth et al. (2008; 2009) scheme, we have classified 9 Class I and 75 Class II objects from mid-IR data. Other candidate YSOs are identified from near-IR $(J-H)/(H-K)$ CC diagram. We have identified 152 candidate YSOs from IR excess and $H_\alpha$ emission towards the region.

\item We characterized these YSOs from various colour-magnitude diagrams. From $V/(V-I)$ CMD, we have estimated the cluster age which is in the range of $\sim$ 2$-$5 Myr. From mid-IR data, we have estimated the disc fraction of  $\sim$ 58\%, which corresponds to an age of $\sim$ 2-5 Myr. The masses of the candidate YSOs are found to be in the range  $\sim$ 0.1 to 2.0 M$_\odot$ in the  $J/(J-H)$ CMD.

\item The morphology of the region has been studied from spatial distribution of YSOs, stellar density distribution, signature of dust in various optical-infrared images along with the extinction map.

\end{enumerate}

\section*{Acknowledgments}
 We thank the anonymous referee for valuable comments which further improved the quality of the paper. This research work is financially supported by S N Bose National Centre for Basic Sciences under Department of Science and Technology, Govt. of India.   This publication makes use of data from The UKIRT Infrared Deep Sky Survey or UKIDSS which is a next generation near-infrared sky survey using the wide field camera (WFCAM) on the United Kingdom Infrared Telescope on Mauna Kea in Hawaii. We also use data from Two-micron All-Sky Survey (2MASS), which is a joint project of the University of Massachusetts and Infrared Processing and Analysis Center/California Institute of Technology, funded by National Aeronautics and Space Administration and the National Science Foundation. This paper uses data observations made with the Spitzer Space Telescope, which is operated by the Jet Propulsion Laboratory, California Institute of Technology, under contract with NASA. The authors are thankful to the HTAC members and staff of HCT, operated by Indian Institute of Astrophysics (Bangalore); JTAC members and staff of 1.04m Sampurnanand telescope operated by Aryabhatta Research Institute of Observational Sciences ( Nainital).

\subsection{Subsection title}

\bsp

\label{lastpage}


\begin{thebibliography}{99}

\bibitem[\protect\citeauthoryear{AllenLE}{2004}]{b1} Allen L. E., Calvet, N., D'Alessio, P., Merin, B., Hartmann, L., Megeath, S. T., Gutermuth, R. A., Muzerolle, J., Pipher, J. L., Myers, P. C., Fazio, G. G., 2004, ApJS, 154, 363
\bibitem[\protect\citeauthoryear{AllenLE}{1995}]{b2} Allen L. E., Storm K. M. 1995, AJ, 109, 3
\bibitem[\protect\citeauthoryear{AllenTS}{1995}]{b3} Allen T. S. et al. 2008, ApJ, 675, 491
\bibitem[\protect\citeauthoryear{Alexander}{2013}]{b4} Alexander, M. J., Kobulnicky, H. A., Kerton, C. R., \& Arvidsson, K., 2013, ApJ, 770, 1
\bibitem[\protect\citeauthoryear{Avedisova}{1984}]{b5} Avedisova, V. S., \& Kondratenko, G. I., 1984, Nauchnye Informasii, 56, 59
\bibitem[\protect\citeauthoryear{Barentsen}{2011}]{b6} Barentsen et al. 2011, MNRAS, 415, 103
\bibitem[\protect\citeauthoryear{Barentsen}{2014}]{b7} Barentsen et al. 2014, MNRAS, 444, 3257
\bibitem[\protect\citeauthoryear{Bessell}{1988}]{b8} Bessell M. \& Brett J. M., 1988, PASP, 100, 1860
\bibitem[\protect\citeauthoryear{Blitz}{1982}]{b9} Blitz, L., Fich, M., \& Stark, A. A. 1982, ApJS, 49, 183
\bibitem[\protect\citeauthoryear{Bressan}{2012}]{b10} Bressan A., Marigo P., Girardi L., Bernado S., Cero C. D., Rubele S. \& Nanni A., 2012, MNRAS, 427, 127 
\bibitem[\protect\citeauthoryear{Briceno}{2002}]{b11} Briceno, C., Luhman, K. L., Hartmann, L., Stauffer, J. R., \& Kirkpatrick, J. D., 2002, ApJ, 580, 317
\bibitem[\protect\citeauthoryear{Carpenter}{2001}]{b12} Carpenter J. M., 2001, AJ, 121, 2851
\bibitem[\protect\citeauthoryear{Casertano}{1985}]{b13} Casertano, S. \& Hut, P., 1985, ApJ, 298, 80
\bibitem[\protect\citeauthoryear{Chini}{1984}]{b14} Chini, R., Mezger, P. G., Kreyna, E., \& Gem$\ddot{u}$nd, H. P., 1984, A\&A, 135, L14
\bibitem[\protect\citeauthoryear{Cohen}{1981}]{b15} Cohen J. G., Frogel J. A., Persson S. E., Ellias J. H., 1981, ApJ, 249, 481
\bibitem[\protect\citeauthoryear{Corradi}{2008}]{b16} Corradi et al. 2008, A\&A 480, 409
\bibitem[\protect\citeauthoryear{Cutri}{2003}]{b17} Cutri R. M., et al., 2003, The IRSA 2MASS All Sky Point Source Catalog, NASA/IPAC Infrared Science Archive, http://irsa.ipac.caltech.edu/applications/Gator
\bibitem[\protect\citeauthoryear{Dahm}{2005}]{b18} Dahm S., 2005, AJ, 130, 1805
\bibitem[\protect\citeauthoryear{Drew}{2005}]{b19} Drew et al. 2005, MNRAS, 362, 753
\bibitem[\protect\citeauthoryear{Elias}{1982}]{b20} Elias et al. 1982, AJ, 87, 1029
\bibitem[\protect\citeauthoryear{Evans}{2009}]{b21} Evans, II, N. J., Dunham, M. M., Jørgensen, J. K., et al. 2009, ApJS, 181, 321
\bibitem[\protect\citeauthoryear{Flaherty}{2007}]{b22} Flaherty K. M., Pipher J. L., Megeath S. T., Winston E. M., Gutermuth R. A., Muzerolle J., Allen L. E., Fazio G. G., 2007, ApJ, 663, 1069
\bibitem[\protect\citeauthoryear{Girardi}{2002}]{b23} Girardi., Bertelli G., Chiosi C., Groenewegen M. A. T., Marigo P., Salasnich B., Weiss A., 2002, A\&A, 391, 195
\bibitem[\protect\citeauthoryear{Gonz\'{a}lez}{2005}]{b24} Gonz\'{a}lez-Solares E. A. et al. 2008, MNRAS, 388,89
\bibitem[\protect\citeauthoryear{Gutermuth}{2005}]{b25} Gutermuth R. A., Megeath S. T., Pipher J. L., Williams J. P., Allen L. E., Myers P. C., Raines S. N. 2005, ApJ, 632, 397
\bibitem[\protect\citeauthoryear{Gutermuth}{2008}]{b26} Gutermuth R. A., Myers P. C., Megeath S. T., Allen L. E., Pipher J. L., Muzerolle J., Porras A., Winston E., Fazio G., 2008, ApJ, 674, 307
\bibitem[\protect\citeauthoryear{Gutermuth}{2009}]{b27} Gutermuth R. A., Megeath S. T., Myers P. C., Allen L. E., Pipher J. L., Fazio G. G., 2009, ApJs, 184, 18	
\bibitem[\protect\citeauthoryear{Haisch}{2001}]{b28} Haisch, K. E. Jr., Lada, E. A. \& Lada, C. J., 2001, ApJ, 553, L153
\bibitem[\protect\citeauthoryear{Hern\'{a}ndez}{2007}]{b29} Hern\'{a}ndez, J., Hartmann, L.,Megeath, T., et al. 2007, ApJ, 662, 1067
\bibitem[\protect\citeauthoryear{Herczeg}{2014}]{b30} Herczeg, G. J., \& Hillenbrand, L. A., 2014, ApJ, 786, 97
\bibitem[\protect\citeauthoryear{Hillenbrand}{1997}]{b31} Hillenbrand, L. A. 1997, AJ, 113, 1733
\bibitem[\protect\citeauthoryear{Hillenbrand}{2002}]{b32} Hillenbrand, L. A. 2002, arXiv: astro-ph/0210520
\bibitem[\protect\citeauthoryear{Horner}{1997}]{b33} Horner D. J., Lada E. A., Lada C. J., 1997, AJ, 113, 5
\bibitem[\protect\citeauthoryear{Jacoby}{1984}]{b34} Jacoby G. H., Hunter D. A., Christian C. A., 1984, ApJS, 56, 257
\bibitem[\protect\citeauthoryear{Jose}{2012}]{b35} Jose J., Pandey A. K., Ogura K., Samal. M. R., Ojha D. K., Bhatt B. C., Chauhan N., Eswaraiah C., Mito H., Kobayashi N., Yadav R. K., 2012, MNRAS, 424, 2486
\bibitem[\protect\citeauthoryear{Jose}{2013}]{b36} Jose J., Pandey A. K., Samal M. R., Ojha D. K., Ogura K., Kim J. S., Kobayashi N., Goyal A., Chauhan N., Eswaraiah C., 2013, MNRAS, 432, 3445
\bibitem[\protect\citeauthoryear{Kenyon}{1995}]{b37} Kenyon S., \& Hartman 1995, ApJS, 101, 117
\bibitem[\protect\citeauthoryear{Kislyakov}{1995}]{b38} Kislyakov, A. G., \& Turner B. E. 1995, AZh, 72, 168
\bibitem[\protect\citeauthoryear{Koorneef}{1983}]{b39} Koorneef J., 1983, A\&A, 128, 84
\bibitem[\protect\citeauthoryear{Kunter}{1980}]{b40} Kutner, M. L., Machnik, D. E., Tucker, K. D., \& Dickman, R. L., 1980, ApJ, 237, 734
\bibitem[\protect\citeauthoryear{Lada \& Adams}{1992}]{b41} Lada C. J., Adams F. C., 1992, ApJ, 393, 278
\bibitem[\protect\citeauthoryear{Lada \& Lada}{1995}]{b42} Lada, E. A., \& Lada, C. J., 1995, AJ, 109, 1682
\bibitem[\protect\citeauthoryear{Lada \& Lada}{2003}]{b43} Lada C. J., \& Lada E. A., 2003, ARA\&A, 41, 57
\bibitem[\protect\citeauthoryear{Lada}{1994}]{b44} Lada C. J., \& Lada E. A., Cliemens D. P., Bally J., 1994, ApJ 429, 694
\bibitem[\protect\citeauthoryear{Lada}{2010}]{b45} Lada C. J., Lombardi M., Alves J. F., 2010, ApJ, 724, 687
\bibitem[\protect\citeauthoryear{Lada}{1991}]{b46} Lada C. J., Evans, N. J., Depoy, D. L., \& Gatley, I., 1991, ApJ, 371, 171
\bibitem[\protect\citeauthoryear{Landolt}{1992}]{b47} Landolt A. U., 1992. AJ, 104, 340
\bibitem[\protect\citeauthoryear{Leisawitz}{1989}]{b48} Leisawitz D., Bash F. N. \& Thaddeus P., 1989, ApJS, 70, 731
\bibitem[\protect\citeauthoryear{Lundquist}{2014}]{b49} Lundquist M. J., Kobulnicky H. A., Alexander M. J., Kerton C. R., Arvidsson K., 2014, ApJ, 784, 111
\bibitem[\protect\citeauthoryear{Lawrence}{2007}]{b50}Lawrence et al. 2007; MNRAS, 379, 1599
\bibitem[\protect\citeauthoryear{Lucas}{2008}]{b51} Lucas et al. 2008, MNRAS, 391, 136
\bibitem[\protect\citeauthoryear{Luhman}{2004}]{b52} Luhman, K. L., Peterson, D. E., Megeath, S. T., 2004, ApJ, 617, 565 
\bibitem[\protect\citeauthoryear{Meyer}{1997}]{b53} Meyer M., Calvet N., Hillenbrand L. A., 1997, AJ, 114, 288
\bibitem[\protect\citeauthoryear{Megeath}{2004}]{b54} Megeath, S. T., Gutermuth, R. A., Allen, L. E., Pipher, J. L., Myers, P. C., \& Fazio, G. G., 2004, ApJS, 154, 367
\bibitem[\protect\citeauthoryear{Ojha}{2004}]{b55} Ojha D. K. et al., 2004, ApJ, 608, 797
	\bibitem[\protect\citeauthoryear{Oke}{1990}]{b56} Oke J. B. 1990, AJ, 99, 5
\bibitem[\protect\citeauthoryear{Pecaut}{2004}]{b57}	Pecaut M. J. \&  Mamajek, E. E., 2013, ApJS, 208, 9
\bibitem[\protect\citeauthoryear{Persi}{1994}]{b58} Persi, P., Roth, M., Tapia, M., Ferrari-Toniolo M., \& Marenzi, A. R., 1994, A\&A, 282, 474
\bibitem[\protect\citeauthoryear{Petrossian}{1985}]{b59} Petrossian, V. M., 1985, ATR, 22, 423
\bibitem[\protect\citeauthoryear{Prabhu}{2014}]{b60} Prabhu T.P. 2014, Proc. of Ind. Nat. Sc. Acad., 80, 887
\bibitem[\protect\citeauthoryear{Rebull}{2010}]{b61} Rebull, L. M., Padgett, D. L., McCabe, C.-E., et al. 2010, ApJS, 186, 259
\bibitem[\protect\citeauthoryear{Racine}{1968}]{b62} Racine, R., 1968, AJ, 73, 233
\bibitem[\protect\citeauthoryear{Rice}{2012}]{b63} Rice, T., Wolk, S., Aspin, C., 2012, ApJ, 755, 65
\bibitem[\protect\citeauthoryear{sagar}{1999}]{b64} Sagar, R., ``Some new initiatives in optical astronomy at UPSO, Nainital'', 1999, in Current Science, 77, 643-652
\bibitem[\protect\citeauthoryear{Samal}{2015}]{b65} Samal M. R., Ojha D. K., Jose J., Zavagno A., Takahashi S., 
Niechel B., Kim J. S., Chauhan N., Pandey A. K., Zinchenko I., Tamura M., Ghosh S. K., 2015, A\&A, 581, 5S
\bibitem[\protect\citeauthoryear{Samal}{2014}]{b66} Samal M. R., Zavagno A., Deharveng L., Molinari S., Ojha D. K., Paradis D., Tig\'{e} J., Pandey A. K., Russeil D., 2014, A\&A 566, A122
\bibitem[\protect\citeauthoryear{Schmidt-kaler}{1982}]{b67} Schmidt-Kaler Th., 1982, in Schaifers K., Voigt H. H., Landolt H., eds, Landolt-Bornstein, Vol. 2b, Springer, Berlin, p. 19
\bibitem[\protect\citeauthoryear{Stetson}{1987}]{b68} Stetson, P. B., 1987, PASP, 99, 191
\bibitem[\protect\citeauthoryear{Stetson}{1992}]{b69} Stetson, P. B., 1992, in Astronomical Society of the Pacific Conference Series, Vol. 25, Astronomical Data Analysis
\bibitem[\protect\citeauthoryear{Strom}{1989}]{b70} Strom, K. M., Strom, S. E., Edwards, S., Cabrit S., Skrutskie, M. F., 1989, AJ, 97, 1451
\bibitem[\protect\citeauthoryear{Tapia}{1997}]{b71} Tapia, M., Persi, P., Bohigas, J., \& Ferrari-Toniolo, M., 1997, AJ, 113, 1769
\bibitem[\protect\citeauthoryear{Torres-Dodgen}{1993}]{b72} Torres-Dodgen A. V., Weaver W. B., 1993, PASP, 105, 693
\bibitem[\protect\citeauthoryear{Van}{1966}]{b73} Van der Berg, S., 1966, AJ, 71, 990
\bibitem[\protect\citeauthoryear{Wachmann}{1996}]{b74} Wachmann A. A., 1996, Astr. Abh. Stern. Hamburg-Bergedorf, 7, 341
\bibitem[\protect\citeauthoryear{Walborn}{1990}]{b75} Walborn N. R., Fitzpatrick E. L., 1990, PASP, 102, 379



\end{thebibliography}
\end{document}